\documentclass[12pt]{article}
\usepackage[T1]{fontenc}
\usepackage[utf8]{inputenc}
\usepackage[normalem]{ulem}

 \usepackage{color}
\usepackage{bm}
\usepackage{slashed}
\usepackage{varioref}
\usepackage[usenames,dvipsnames]{xcolor}
\usepackage{graphicx}
\usepackage{amsmath}
\usepackage{amssymb}
\usepackage{amsfonts}
\usepackage{amsthm}
\usepackage{array}
\usepackage{yfonts}
\usepackage{float}
\usepackage{comment}
\usepackage{caption}
\usepackage[sort]{cite}
\usepackage{subcaption}
\usepackage{wrapfig}
\usepackage[
      colorlinks=true,
      linkcolor=blue,
      urlcolor=blue,
      filecolor=blue,
      citecolor=red,
      pdfstartview=FitV,
      pdftitle={},
      pdfauthor={},
      pdfsubject={},
      pdfkeywords={},
      bookmarksopen=true
]{hyperref}

\usepackage{epsfig}

\usepackage{hyperref}
\newcommand{\mcN}{{\mycal N}}

\newcommand{\sectionofScri}%
{{ \,\,\,\,\mathring{\!\!\!\!\mcN}}}

\newcommand{\eeal}[1]{\label{#1}\end{eqnarray}}

\DeclareFontFamily{OT1}{rsfs}{}
\DeclareFontShape{OT1}{rsfs}{m}{n}{ <-7> rsfs5 <7-10> rsfs7 <10-> rsfs10}{}
\DeclareMathAlphabet{\mycal}{OT1}{rsfs}{m}{n}

\newcommand{\mmr}[1]{\mnotex{{\bf mm:} {\color{violet} #1}}}

\definecolor{applegreen}{rgb}{0.55, 0.71, 0.0}
\definecolor{armygreen}{rgb}{0.29, 0.33, 0.13}

\definecolor{caribbeangreen}{rgb}{0.0, 0.8, 0.6}

\newcounter{mnotecount}[section]

\renewcommand{\themnotecount}{\thesection.\arabic{mnotecount}}

\newcommand{\mnotex}[1]
{\protect{\stepcounter{mnotecount}}$^{\mbox{\footnotesize
        $
        \bullet$\themnotecount}}$ \marginpar{
    \raggedright\tiny\em
    $\!\!\!\!\!\!\,\bullet$\themnotecount: #1} }

\newcommand{\ptc}[1]{\mnotex{{\bf ptc:} {  #1}}}
\newcommand{\ptcrr}[1]{\mnotex{{\bf ptc:} {  #1}}}

\newcommand{\bel}[1]{\begin{equation}\label{#1}}

\newcommand{\eeq}{\end{equation}}
\newcommand{\ee}{\end{equation}}
\newcommand{\beqa}{\begin{eqnarray}}
\newcommand{\eeqa}{\end{eqnarray}}
\newcommand{\beqan}{\begin{eqnarray*}}
\newcommand{\eeqan}{\end{eqnarray*}}
\newcommand{\ba}{\begin{array}}
\newcommand{\ea}{\end{array}}

\newcommand{\ptcheck}[1]{\ptc{checked on #1}}

\def\beq{\begin{eqnarray}}
\def\eeq{\end{eqnarray}}

\def\be{\begin{equation}}
\def\ee{\end{equation}}
\def\bea{\begin{eqnarray}}
\def\eea{\end{eqnarray}}

{\catcode `\@=11 \global\let\AddToReset=\@addtoreset}
\AddToReset{equation}{section}

\newcommand{\jhbr}[1]{\mnotex{{\bf jh:} {\color{blue} #1}}}

\newcommand{\tscomr}[1]{\mnotex{{\bf ts:} {  #1}}}

{\catcode `\@=11 \global\let\AddToReset=\@addtoreset}
\AddToReset{equation}{section}

\renewcommand{\ptcrr}[1]{}

\renewcommand{\mmr}[1]{}
\renewcommand{\tscomr}[1]{}

\renewcommand{\ptcheck}[1]{}
\renewcommand{\jhbr}[1]{}

\begin{document}

\title{\bf{Cosmological perturbation theory\\ of primordial compact sources}
}
\date{}

\maketitle

\vspace{-1.5cm}

\centerline{\large{\bf{Geoffrey 
Comp\`ere}$^{a}$\footnote{geoffrey.compere@ulb.be},  Sk Jahanur 
Hoque$^{a,b,c}$\footnote{jahanur.hoque@hyderabad.bits-pilani.ac.in}
}}
\vspace{6pt}

\bigskip\medskip
\centerline{\textit{{}$^{a}$ Universit\'e Libre de Bruxelles, International Solvay Institutes}}
\centerline{\textit{and Brussels Laboratory of the Universe (BLU-ULB), CP 231, B-1050 Brussels, Belgium}}

\medskip
\centerline{\textit{{}$^{b}$ Birla Institute of Technology and Science, Pilani, Hyderabad Campus, }}
\centerline{\textit{Jawaharnagar, Hyderabad 500 078, India}}

\medskip
\centerline{\textit{{}$^{c}$ Institute of Theoretical Physics, Faculty of Mathematics and Physics, Charles University, }}
\centerline{\textit{V Holešovičkách 2, 180 00 Prague 8, Czech Republic}}

\vspace{1cm}


\begin{abstract}
We construct a position-space cosmological perturbation theory around spatially flat Friedmann-Lema\^itre-Robertson-Walker geometries that allows to model localized primordial  sources of gravitational waves. The equations of motion are decoupled using a generalized harmonic gauge, which avoids the use of a scalar-vector-tensor decomposition. We point out that  sources cannot generically be defined in a compact domain due to fluctuations of the cosmic perfect fluid. For power law cosmologies, we obtain the exact Green's function necessary to solve for all metric perturbations in terms of a hypergeometric function, which matches with a Green's function derived earlier by Chu. This allows us to derive the closed form expression of the linearized metric perturbation generated by sources up to quadrupolar order in the multipolar expansion.
\end{abstract}
\newpage
\tableofcontents

\section{Introduction}

Gravitational waves (GW) propagated through a Friedmann-Lema\^itre-Robertson-Walker (FRLW) universe are well understood when the geometric optics approximation is valid. Denoting the GW frequency of the source at emission as $f_\text{gw}$ and the conformal time at emission as $\eta_\text{emiss}$ the geometric optics approximation reads as\footnote{The scale of the FRLW background curvature is proportional to $(\partial_\eta a/a)^2$ and $\partial_\eta^2  a/a$ which both scale as $1/\eta^2$ at early times in the $\Lambda$CDM model, see Section \ref{sec:eqs}.}
\begin{align}
2 \pi f_\text{gw} \gg \frac{1}{\eta_\text{emiss}}. \label{GeomOpt}
\end{align}
For all sources that obey this condition, the effect of the cosmological background on gravitational wave propagation simply amounts to a redshift of the GW frequency, the modulation of the GW amplitude by the scale factor and a redshift of the masses of the source \cite{Maggiore:2007ulw, Deruelle:1984hq}.

Now, for sources at sufficiently early times $\eta_\text{emiss} \to 0$, the geometric optics approximation breaks down, and a fully relativistic analysis of perturbations is required\footnote{The geometric optics approximation is also violated near compact objects when $2 \pi f_\text{gw} \gg \frac{1}{L_\text{c.o.}}$ where $1/L_{c.o.}^2$ is the curvature scale of the compact object. However, in that context, cosmological scales are negligible and ordinary Post-Newtonian methods can be applied.}. Such primordial perturbations produce GW fluctuations at cosmological scales, which are well beyond the reach of current GW detectors, but which indirectly lead to electromagnetic signatures, such as in the cosmic microwave background (CMB) spectrum.


The main formalism currently used for the computation of perturbations of FRLW spacetimes is based on the mathematically elegant SVT decomposition, which allows to fully decouple scalar, vector and tensor modes and write down gauge invariant cosmological perturbations   \cite{Bardeen:1980kt,Kodama:1984ziu,Durrer:1993tti}.  When combined with inflation and quantum mechanics, cosmological perturbation theory allows to infer the large scale structure of the current universe \cite{Mukhanov:1990me}. 

Due to the spatial non-locality of the SVT decomposition, perturbations are typically expanded in Fourier modes. Fourier domain is convenient for the study of the fundamental stochastic GW background as the spectrum of perturbations generated during inflation is nearly scale invariant \cite{Planck:2015sxf}. However, when considering individual localized sources, the SVT formalism requires to solve Poisson equations in order to  reconstruct the metric perturbation, which is not straightforward.   It is convenient instead to derive the perturbation equations in position space as most of the tools derived in post-Newtonian methods in the non-cosmological context are precisely derived in position space using a basis of symmetric trace-free (STF) tensors \cite{Damour:1990gj}. In order to fill this gap, we derive in this paper a formalism that solves for the propagation of GW in FRLW universe without using neither the geometric optics approximation nor the SVT decomposition. Instead, the equations will be decoupled in a basis of symmetric trace-free (STF) tensors by introducing a generalized harmonic gauge. Our formalism is therefore adapted to the study of compact sources in the early universe, such as \emph{e.g.} cosmic strings or primordial black hole mergers \cite{Sasaki:2018dmp}. The generalized harmonic gauge that we define is the first gauge to our knowledge that allows to fully decouple the linearized perturbations without using a SVT decomposition, which departs from standard treatments of cosmological perturbation theory \cite{Baumann:2009ds}.

Position space methods are particularly convenient to study memory effects. In the cosmological context, for sources that obey the geometric optics approximation, the memory is simply increased by a redshift factor as compared with the non-cosmological context \cite{Tolish:2016ggo, Bieri:2015jwa, Bieri:2017vni}. In the absence of the geometric optics approximation, memory effects have been derived by first deriving the Green's function for metric perturbations, and solving for the transverse traceless modes, see \cite{Chu:2016qxp,Chu:2015yua,Compere:2023ktn} for de Sitter, \cite{Chu:2016ngc} for power-law cosmologies and \cite{Jokela:2023suz,Jokela:2023ytv} for the  $\Lambda$CDM model. Yet, the literature on Green's functions for FRLW cosmologies is bewildering, as several authors find different results \cite{Chu:2015yua,Chu:2016qxp,Caldwell:1993xw,Haas:2004kw}. One aspect of our work is to rederive the relevant Green's functions from first principles and identify how it compares with the literature. 

One important question in cosmology is the very definition of a compact source. We emphasize in our treatment that defining a vanishing stress-energy tensor perturbation outside a compact support is fundamentally inconsistent with the local conservation of the stress-energy tensor. This will lead us to define a nearly spatially compact source, which still admits a non-compact energy density perturbation. We will finally derive the explicit metric perturbations for a nearly compact source in power law cosmologies. We will follow the consistent quadrupolar truncation scheme introduced in \cite{Compere:2023ktn}. 

The rest of the paper is organized as follows. We decouple the perturbation equations in Section \ref{sec:eqs} and comment upon the de Sitter background, power law cosmologies and the $\Lambda$CDM model. The relevant Green's functions for power-law cosmologies are derived in Section \ref{sec:Green} and compared with the literature. The consistent quadrupolar truncation is described in Section \ref{sec:multi}. In Section \ref{sec:quad} we compute the exact linearized metric perturbation around an arbitrary power-law cosmology background at quadrupolar order. We conclude in Section \ref{sec:ccl}.

\section{Decoupled perturbations in generalized harmonic gauge}
\label{sec:eqs}

Linear perturbation theory can be defined covariantly as follows. We consider a one-parameter family of metrics, $g_{\mu \nu}(\lambda)$ which is differentiable with respect to $\lambda$ at $\lambda =0$. The background metric is  $\bar{g}_{\mu \nu}:=g_{\mu \nu}|_{\lambda =0}$ while linearized perturbations belong to the tangent space at $\lambda=0$, $h_{\mu \nu}:=\frac{d g_{\mu \nu}(\lambda)}{d\lambda}|_{\lambda =0}$.  In order to write down the equations of motion, we choose a coordinate system and write $g_{\mu \nu}(\lambda, x)=\bar{g}_{\mu \nu}(x)+\lambda h_{\mu \nu }(x)+O(\lambda^2)$. The linearized equations are obtained by substituting the metric expansion in Einstein's equations $G_{\mu \nu}+{\Lambda g_{\mu \nu} }= 8 \pi G T_{\mu \nu}$ and keeping terms of order $\lambda$.

The spatially flat FLRW metric in conformal coordinates $x^\mu = (\eta, x^i)$ has the form 
\bea \label{21IX25.01}
\bar{g}_{\mu \nu} dx^{\mu} dx^\nu= a^{2}(\eta)(-d\eta^{2}+\delta_{ij} dx^{i} dx^{j}).
\eea
Conformal time $\eta$, and 
cosmological proper time $t$ are related as $a^{2} d\eta^{2}=dt^{2}$. 
Hence $\partial_{\eta}=a\partial_{t}$. The deceleration parameter is defined as 
\bea 
q:=-\frac{a\partial_{t}^{2}a}{(\partial_{t}a)^{2}} = 1-\frac{a(\eta)\ddot a(\eta)}{\dot a(\eta)^2},
\eea
where $\dot a=\partial_\eta a$. FRLW spacetime is a solution to the background Einstein's equations $\bar{G}_{\mu \nu}+ {\Lambda \bar{g}_{\mu \nu}}=8 \pi G \bar{T}_{\mu \nu}$ where the stress-tensor is the one of a perfect fluid at rest
\begin{align}\label{barTmunu}
\bar T_{\mu \nu}=(\epsilon+p) U_{\mu}U_{ \nu}+p\bar{g}_{\mu \nu}, \qquad U^\mu = \frac{1}{a(t)} \delta^\mu_{\eta}.
\end{align}
Equivalently, $\bar T^{\mu}{}_{\nu}=\mbox{diag}(-\epsilon,p,p,p)$. Einstein's equations and the conservation of the background stress-energy tensor reduce to the FRLW equations
\bea \label{19XII25.01}
\bigg(\frac{\dot a}{a}\bigg)^{2}-\frac{\Lambda }{3}a^{2} &=& \frac{8\pi G}{3} a^2 \epsilon ,\\
q\bigg(\frac{\dot a}{a}\bigg)^{2}+  \frac{\Lambda}{3} a^{2}&=&\frac{4\pi G}{3} a^{2}(\epsilon+3p),
\eea
and to the continuity equation  
\begin{align}
  \dot \epsilon+\frac{3\dot a}{a}(\epsilon+p)=0.  
\end{align}

The linearized Einstein equations read as 
\bea \label{10XI25.01}
R_{\mu\nu}^{(1)}-\frac{1}{2}(\bar{g}_{\mu \nu} R^{(1)}+\bar{R} h_{\mu \nu}-\bar{g}_{\mu \nu} \bar{R}_{\alpha \beta} h^{\alpha \beta})+\Lambda h_{\mu \nu}= 8 \pi G \delta T_{\mu \nu}.
\eea
Using the expressions of the linearized Ricci tensor $R^{(1)}_{\mu\nu}$ and scalar $R^{(1)}$, Eq. \eqref{10XI25.01} becomes
\bea \nonumber
&&-\frac{1}{2} \bar{\square} h_{\mu \nu} -\frac{1}{2}(\bar{\nabla}_{\mu 
}\bar{\nabla}_{\nu}-\bar{g}_{\mu \nu} \bar{\square})h-\frac{\bar{R}}{2} 
h_{\mu \nu} +\frac{1}{2} (\bar{\nabla}_{\mu} \bar{\nabla}_{\alpha} 
h^{\alpha}{}_{\nu}+ \bar{\nabla}_{\nu} \bar{\nabla}_{\alpha} h^{\alpha}
{}_{\mu}-\bar{g}_{\mu \nu} \bar{\nabla}_{\alpha} \bar{\nabla}_{\beta} h^{\alpha \beta})\\ \label{6II25.01}
&&+\bar{R}_{\mu \alpha \beta \nu} h^{\alpha \beta} + \frac{1}{2}(\bar{R}_{\mu \alpha} h^{\alpha}{}_{\nu}+\bar{R}_{\nu \alpha} h^{\alpha}{}_{\mu}+ \bar{g}_{\mu \nu} \bar{R}_{\alpha \beta} h^{\alpha \beta}) +{\Lambda h_{\mu \nu} } = 8 \pi G \delta T_{\mu \nu}.
\eea
It is convenient to write the linearized equations of motion in terms of trace-reversed perturbations, $\tilde{h}_{\mu \nu}:= h_{\mu \nu}-\frac{1}{2} \bar{g}_{\mu\nu} h$. Denoting $B_{\mu}:= \bar{\nabla}_{\alpha}\tilde{h}^{\alpha}{}_{\mu}$ and using the trace-reversed perturbations,  the linearized equations become
\begin{align} \nonumber
&\frac{1}{2}(-\bar{\square}\tilde{h}_{\mu \nu}+\bar{\nabla}_{\mu }B_{\nu}+\bar{\nabla}_{\nu }B_{\mu}-\bar{g}_{\mu \nu}(\bar{\nabla}^{\alpha} B_{\alpha})) + \bar{R}_{\mu \alpha \beta \nu} \tilde{h}^{\alpha \beta}-\frac{\bar{R}}{2} \tilde{h}_{\mu \nu} \\ \label{9II25.01}
&+\frac{1}{2} (\bar{R}_{\mu \alpha} \tilde{h}^{\alpha}{}_{\nu}+\bar{R}_{\nu \alpha} \tilde{h}^{\alpha}{}_{\mu} +\bar{g}_{\mu\nu} \bar{R}_{\alpha \beta} \tilde{h}^{\alpha \beta})+ {\Lambda (\tilde{h}_{\mu\nu}-\frac{1}{2}\bar{g}_{\mu\nu} \tilde{h})}
= 8\pi G \delta T_{\mu\nu}.
\end{align}
We now define generalized harmonic gauge as the gauge fixing condition $B_{\mu}= -\frac{2 \dot a}{a^{3}} \tilde{h}_{0\mu}$.  In terms of trace-reversed perturbations, the 
generalized harmonic gauge condition reads as
\bea \label{genH}
\partial^{\alpha} \tilde{h}_{\alpha \mu}-\bigg(\frac{\dot a}{a}\bigg)\delta_{\mu}^{0} \tilde{h}_{\alpha \beta}\eta^{\alpha \beta}=0.
\eea

Substituting the background to the FRLW universe in Eq. \eqref{9II25.01} and using generalized harmonic gauge, the linearized equations on the FLRW background are given by
\begin{align} \nonumber
\square \tilde{h}_{\mu \nu}+ 2\bigg(\frac{\dot a}{a}\bigg)\partial_{0} \tilde{h}_{\mu\nu}-2 \bigg(\frac{\dot a}{a}\bigg)^{2}\bigg \{-q \delta_{\mu}^{0} \delta_{\nu}^{0} \tilde{h}_{\alpha}{}^{\alpha}-(1-3q) \tilde{h}_{\mu \nu}{-q}(\delta_{\mu}^{0}\tilde{h}_{0\nu}+\delta_{\nu}^{0}\tilde{h}_{0\mu})\\ \label{10II25.01}
+\eta_{\mu \nu} \tilde{h}_{00} (1+q)+ \eta_{\mu \nu} \tilde{h}_{\alpha}{}^{{\alpha}}(1-\frac{q}{2})\bigg\}
-{2 \Lambda a^{2}(\tilde{h}_{\mu\nu}-\frac{\eta_{\mu \nu}}{2}\tilde{h}_{\alpha}{}^{\alpha})}=-16\pi G a^{2} \delta T_{\mu \nu},
\end{align}
where $\square$ denotes the usual wave operator around Minkowski, and $\tilde{h}_{\alpha}{}^{\alpha}:=\tilde{h}_{\alpha \beta} \eta^{\alpha \beta}$. 

Let us first specialize to the de Sitter background where $\Lambda=3H^2$ and $a=-\frac{1}{H\eta}$, which implies $(\frac{\dot a}{a})=-\frac{1}{\eta}$ and $q=-1$. In that case, the gauge condition reduces to  $\bar \nabla_\alpha \tilde h^\alpha_{\;\; \mu}= \frac{2\Lambda}{3} \eta \tilde{h}_{0\mu}$, which is identical to the gauge condition defined in \cite{deVega:1998ia}. Around de Sitter, Eq. \eqref{10II25.01} becomes
\bea 
\square \tilde{h}_{\mu \nu}-\frac{2}{\eta} \partial_{0} \tilde{h}_{\mu\nu} -\frac{2}{\eta^{2}} \bigg\{\delta_{\mu}^{0} \delta_{\nu}^{0}\tilde{h}_{\alpha}{}^{\alpha}-\tilde{h}_{\mu\nu} + \delta_{\mu}^{0} \tilde{h}_{0\nu}+\delta_{\nu}^{0} \tilde{h}_{0\mu}\bigg\}=-16\pi G a^{2} \delta T_{\mu\nu}.
\eea
This equation matches exactly with Eq. (69) of \cite{Date:2015kma}, and Eq. (3.23) of \cite{deVega:1998ia} after correctly identifying the field variables. It was shown in \cite{deVega:1998ia} that these linearized equations can be totally decoupled after using an appropriate choice of field variables. 

Let us now use the gauge condition \eqref{genH} around the spatially flat FLRW background and seek how to decouple the equations. 
We introduce the rescaled linear perturbation $\tilde \chi_{\mu\nu} = a^{-1}\tilde h_{\mu\nu}$. The equations become 
\begin{align} \nonumber
\square \tilde\chi_{\mu \nu} - 2\bigg(\frac{\dot a}{a}\bigg)^2\bigg\{-q \delta_{\mu}^{0} \delta_{\nu}^{0} \tilde\chi_{\alpha}{}^{\alpha}+\frac{1}{2}(5q-3 ) \tilde\chi_{\mu \nu}{ -q}(\delta_{\mu}^{0} \tilde\chi_{0\nu}+\delta_{\nu}^{0} \tilde\chi_{0\mu} )\\ \label{10II25.02}
+\eta_{\mu \nu} \tilde\chi_{00} (1+q) + \eta_{\mu \nu} \tilde\chi_{\alpha}{}^{\alpha} (1-\frac{q}{2})\bigg\}
-{2 \Lambda a^2(\tilde\chi_{\mu\nu}-\frac{\eta_{\mu \nu}}{2}\tilde\chi_{\alpha}{}^{\alpha})}=-16\pi G  a \delta T_{\mu \nu}.
\end{align}
The spatial trace-free part $\langle ij \rangle$ and the mixed $0i$ parts decouple: 
\begin{subequations}\label{18IX25.01}
\begin{align}
 \square \tilde\chi_{\langle ij \rangle} +\left( -2\Lambda a^2 + (3-5q)\frac{\dot  a^2}{a^2}\right)\tilde{\chi}_{\langle ij \rangle }  &=-16 \pi G a \delta T_{\langle ij \rangle} ,  \\ 
\square \tilde \chi_{0i} +\left( -2\Lambda a^2 + 3(1-q) \frac{\dot  a^2}{a^2} \right)\tilde \chi_{0i}
&=-16\pi G a \delta T_{0i}.
\end{align}    
\end{subequations}
The scalars $\tilde \chi_{00}$ and $\tilde \chi_{ii}$ remain coupled with one another in general: 
\begin{align}
\square \left( \begin{array}{cc} \tilde \chi_{00} \\ \tilde \chi_{ii}\end{array} \right) +\left( \begin{array}{cc} -\Lambda a^2 + \frac{3\dot a^2}{a^2} & -\Lambda a^2+ \frac{\dot a^2}{a^2}(2+q) \\
-3\Lambda a^2 - \frac{9\dot a^2}{a^2} q  & \Lambda a^2 - \frac{\dot a^2}{a^2}  (3+2q)
\end{array} \right)  \left( \begin{array}{cc} \tilde \chi_{00} \\ \tilde \chi_{ii}\end{array} \right)=-16 \pi G a \left( \begin{array}{cc} \delta T_{00} \\ \delta T_{ii}\end{array}\right).
\end{align}
We will further decouple these equations below. 

The gauge conditions \eqref{genH} in terms of these variables become 
\begin{align} \label{15XII25.01}
 \partial_\eta \tilde \chi_{00}- \partial_j \tilde \chi_{0j}+\frac{\dot a}{a}\tilde \chi_{ii}=0,\qquad   \left( \partial_\eta + \frac{\dot a}{a}\right)  \tilde \chi_{0i} - \partial_j \tilde \chi_{ij}=0.  
\end{align}

We emphasize that unlike standard cosmological perturbation theory, we have 
not performed the scalar, vector, tensor (SVT) decomposition of the field variables, which is spatially non-local on constant time sections. Here, we performed a local decomposition into irreducible representations of $SO(3)$, a STF (symmetric tracefree)  decomposition. The SVT decomposition in cosmological perturbation theory has 
found vast applications to non-localized sources in stochastic waves of 
astrophysical or cosmological origin. Our formalism is suited to solve for 
inhomogeneous solutions in terms of Green's functions, as we will develop in Section \ref{sec:Green}. In particular, one can 
relate the linearized field in terms of the multipolar structure of spatially compact sources, as performed around Minkowski spacetime \cite{Damour:1990gj}.

Let us now study the conservation of the perturbed stress-energy tensor. After using the Einstein equations, the contracted Bianchi identity $\nabla^{\mu} G_{\mu \nu}=0$ can be written as $\nabla_\mu T^{\mu\nu}=0$. At linear order in the perturbation, this constraint reads as 
\begin{align}
\bar{\nabla}^{\mu} \delta T_{\mu \nu}-\bar{g}^{\mu \rho}(\delta \Gamma^{\lambda}{}_{\rho \nu}\bar{T}_{\lambda \mu}+\delta \Gamma^{\lambda}{}_{\rho \mu}\bar{T}_{\lambda \nu})-h^{\mu \rho}\bar{\nabla}_{\rho} \bar{T}_{\mu \nu}=0. \label{eq:con}
\end{align}
For a background de Sitter or Minkowski space-time, $\bar{T}_{\mu \nu}=0$, and the linearized stress-energy tensor is conserved with respect to background metric. However, for a generic background, the linearized stress-energy is not conserved on its own. In generalized harmonic gauge, Eq. \eqref{eq:con} can be expressed as
\begin{subequations}\label{tildecons}
\begin{align}
\partial_\eta(a\delta  \tilde T_{00}) - \partial_i (a \delta \tilde T_{0i})+\dot a \delta \tilde T_{kk} &=-\frac{1}{4}\partial_\eta( a^2(p+\epsilon) ) (\tilde \chi_{00}+\tilde \chi_{kk} ),\label{cons00shifted} \\ 
(\partial_\eta+\frac{\dot a}{a})(a \delta \tilde T_{0i})-\partial_j (a \delta \tilde T_{i j}) &=0,   
\end{align}    
\end{subequations}
where we defined the shifted stress-energy tensor components 
\begin{subequations}\label{tildeT}
\begin{align}
\delta \tilde T_{00} & \equiv \delta T_{00}+ a\frac{\epsilon-p}{4} (\tilde \chi_{00}+\tilde \chi_{kk}),\\
\delta \tilde T_{0i} & \equiv \delta  T_{0i} - p a\tilde \chi_{0i}, \\ 
\delta \tilde T_{ij} & \equiv \delta T_{ij}- p a\tilde \chi_{ij} - \delta_{ij} a\bigg( \frac{\epsilon-p}{4}\tilde \chi_{kk}+\frac{\epsilon+3p}{4}\tilde \chi_{00} \bigg). 
\end{align}    
\end{subequations}
Note that $ 
\delta \tilde T_{\langle ij \rangle} = \delta T_{\langle ij \rangle} - pa \tilde {\chi}_{\langle ij \rangle}$. 

Let us first discuss the gauge transformation of the shifted stress-energy tensor under an arbitrary linearized diffeomorphism $x^{\mu} \to x^{\mu}-\lambda \xi^{\mu}$. Using $\delta_\xi h_{\mu\nu}=\mathcal L_\xi \bar g_{\mu\nu}$ and $\delta_\xi \delta T_{\mu\nu}=\mathcal L_\xi \bar T_{\mu\nu}$, the components of shifted stress-energy tensor transform as
\bea 
\delta_{\xi} (\delta \tilde T_{00})&=& a^{2} (\epsilon +p) (\dot \xi^{0}-2aH \xi^{0}),\\
\delta_{\xi} ( \delta \tilde T_{0i})&=& a^{2}(\epsilon +p)\partial_{i} \xi^{0},\\
\delta_{\xi} ( \delta \tilde T_{ij})&=& \delta_{ij} \big(a^{2} \xi^{0} \dot{p}+a^{2}(\epsilon+p)(\dot{\xi}^{0}+Ha \xi^{0})\big).
\eea
While the shifted stress-energy tensor is not gauge invariant, its tracefree part $\delta \tilde T_{\langle ij \rangle}$ is gauge invariant, and therefore a priori physical. 

Importantly, one cannot absorb the right-hand side of Eq. \eqref{cons00shifted} by performing a local shift of the stress-energy tensor. Therefore, in the presence of a varying cosmological fluid variable $a^2(p+\epsilon)$, the perturbed stress-energy tensor is not conserved on its own. Using our definition of the shifted stress-energy tensor, this non-conservation can be confined to the $\eta$ component of the shifted stress-energy conservation equation, and be can limited to a term proportional to $\partial_\eta (a^2(p+\epsilon))$ but that term can never been totally removed. We choose to define $\delta \tilde T_{00}$ such that right-hand side is proportional to  $\partial_\eta (a^2(p+\epsilon))$. It is a physical effect of non-conservation of the local stress-energy tensor due to the evolution of the background cosmological fluid. 

We conclude that in FRLW cosmology, we cannot define an isolated stress-tensor perturbation as $\delta \tilde T_{\mu\nu}(x)=0$ for $x$ outside a localized volume $V$ where the source would be localized. That definition would violate the conservation equation \eqref{cons00shifted} since it would amount to fix $\tilde \chi_{00}+\tilde \chi_{kk}=0$ outside of $V$. However, the gauge is already fixed by the generalized harmonic gauge equations \eqref{15XII25.01} and the condition $\tilde \chi_{00}+\tilde \chi_{kk}=0$ would be a constraint on the metric perturbation that is not imposed by Einstein's equations. The definition of a compact source as $\delta \tilde T_{\mu\nu}(x)=0$ for $x$ outside a localized volume $V$ is therefore inconsistent. 

Instead, the best we can do, is to define a nearly isolated stress-tensor perturbation as  $\delta \tilde T_{0i}(x)=\delta \tilde T_{ij}(x)=0$ for $x$ outside a localized volume $V$ where the source is localized, but we allow $\delta \tilde T_{00}(x) \neq 0$ for $x$ outside of $V$. 

In order to describe such nearly-localized sources, we rewrite the linearized equations in terms of the shifted stress-energy tensor. Using the background FRLW equations \eqref{19XII25.01}, we find that the equations totally decouple can be elegantly written as 
\begin{subequations}\label{18IX25.01tilde}
\begin{align}
\mathcal O^- \tilde{\chi}_{\langle ij \rangle} &=-16 \pi G a \delta \tilde{T}_{\langle ij \rangle},\\
\mathcal O^+ \tilde{\chi}_{0i} &=-16 \pi G a \delta \tilde{T}_{0i},\\
\mathcal O^+  (\tilde \chi_{00}+\tilde \chi_{ii}) &= -16 \pi G a (\delta \tilde{T}_{00}+\delta \tilde T_{ii}),\\ 
\mathcal O^- \tilde \chi_{ii} &= -16 \pi G a \delta \tilde T_{ii}.
\end{align}    
\end{subequations}
where the two operators $\mathcal O^\pm$ are defined as 
\begin{align}
\mathcal O^\pm \equiv  \square +\bigg(\frac{\dot a }{a}\bigg)^{2} (1 \pm q)   .
\end{align}
This decoupling of the perturbation equations in the presence of sources around an arbitrary FRLW background, without the use of a SVT decomposition, is one of the main results of this paper.

Let us now discuss special cases of FRLW backgrounds: 

\begin{enumerate}
    \item For de Sitter backgrounds where  $(\frac{\dot a}{a})=-\frac{1}{\eta}$ and $q=-1$, the equations reduce to 
\begin{subequations}\label{dSeqs}
    \begin{align}
(\square +\frac{2}{\eta^2}) \tilde   \chi_{\langle ij \rangle } &= - 16 \pi a \delta T_{\langle ij \rangle},\label{dSeq}\\ 
\square  \tilde  \chi_{0i} &= - 16 \pi a \delta T_{0i}, \\
\square  (\tilde \chi_{00}+\tilde \chi_{ii})  &= - 16 \pi a (\delta T_{00}+\delta T_{ii}), \\
(\square +\frac{2}{\eta^2})\tilde   \chi_{ii}  &= - 16 \pi a \delta T_{ii}. 
\end{align}
\end{subequations}
This reproduces previous results \cite{deVega:1998ia,Ashtekar:2015lxa,Date:2015kma,Compere:2023ktn}. 

\item Power law cosmologies without a cosmological constant admit a scale factor of the form $a(\eta) \sim \vert\eta\vert^\alpha$ which implies $\frac{\dot a}{a}=\frac{\alpha}{\eta}$ and  $q=\frac{1}{\alpha}$. The pressure and density are related as $p = w \epsilon$ with constant $w=(2-\alpha)/(3\alpha)$.  When $\alpha>0$ we take $\eta$ to be positive, $\eta \in [0, \infty)$. When $\alpha<0$ we take $\eta$ to be negative, $\eta \in (-\infty,0]$. Radiation and matter 
dominated universes admit $\alpha=1$ and $2$, respectively. 

The equations of linearized perturbations reduce to 
\begin{subequations}\label{eqspowerlaw}
\begin{align} \label{eqspowerlaw1}
(\square +\frac{\alpha(\alpha-1)}{\eta^2}) \tilde   \chi_{\langle ij \rangle } &= - 16 \pi G a \delta \tilde T_{\langle ij \rangle},\\ \label{eqspowerlaw2}
(\square +\frac{\alpha(\alpha+1)}{\eta^2}) \tilde \chi_{0i} &= - 16 \pi G a \delta \tilde T_{0i}, \\ \label{eqspowerlaw3}
(\square +\frac{\alpha(\alpha+1)}{\eta^2}) (\tilde \chi_{00}+\tilde \chi_{ii})  &= - 16 \pi a (\delta \tilde T_{00}+\delta \tilde T_{ii}), \\ \label{eqspowerlaw4}
(\square +\frac{\alpha(\alpha-1)}{\eta^2})\tilde   \chi_{ii}  &= - 16 \pi G a \delta \tilde T_{ii}. 
\end{align}    
\end{subequations}
Formally, the perturbation equations around de Sitter \eqref{dSeqs} are obtained from these equations with the substitution $\alpha \to -1$ and $\delta \tilde{T}_{\mu\nu} \to \delta T_{\mu\nu}$. 

\item The standard $\Lambda$CDM model is a 6-parameter spatially flat FLRW model with energy density $\epsilon=\epsilon_\text{vac}+\epsilon_m=\Omega_\Lambda \rho_c + \Omega_m \rho_c/a^3$ and pressure $p=-\epsilon_\text{vac}=-\Omega_\Lambda \rho_c$ with $\rho_c=3H_0^2/(8\pi G)$. The Planck 2018 CMB data combined with CMB lensing reconstruction and baryon acoustic oscillation measurements give the best fit values \cite{Planck:2018vyg}: $\Omega_\Lambda=0.69$, $\Omega_m=0.31$ and $H_0=67.7 \text{km s}^{-1}\text{Mpc}^{-1}$. 

In that case the operators $\mathcal O^\pm = \square + V^\pm$ give the potentials 
\begin{align}
 V^+ &= \frac{3}{2}H_0^2 \Omega_m a^{-1}, \\
 V^- &= H_0^2 (\frac{1}{2}\Omega_m a^{-1} +2 \Omega_\Lambda a^2 ),
\end{align}
where we used the background equations \eqref{19XII25.01}.

As also noted in \cite{Jokela:2022rhk}, the scale factor of the $\Lambda$CDM model can be given analytically in terms of the cosmological time $t$ as 
\begin{align}
a(t) = \left( \frac{\Omega_m}{\Omega_\Lambda}\right)^{1/3}\text{sinh}^{\frac 2 3} \left( \frac{3}{2} \sqrt{\Omega_\Lambda} H_0 t \right).    
\end{align}
The function $a(\eta)$ in terms of conformal time is only known numerically as one needs to invert the monotonic relationship
\begin{align}
 H_0 \eta(t) = H_0 \int^t \frac{dt'}{a(t')}=\frac{2}{\Omega_m^{1/2}} \sqrt{a(t)}\;  \mbox{}_2 F_1(\frac{1}{6},\frac 1 2,\frac 7 6,-\frac{\Omega_\Lambda}{\Omega_m}a(t)^3).   
\end{align}
The big bang is at the initial value $\eta=0$. At asymptotically late times, we have 
\begin{align}
H_0 \eta_\text{max}= \frac{2}{(\Omega_m^2 \Omega_\Lambda)^{1/6}}\frac{\Gamma(\frac 1 3)\Gamma (\frac 7 6)}{\sqrt{\pi}} \approx 4.41.     
\end{align}
This is the comoving visibility limit, the largest possible value of $\eta$. Today $a(\eta_\text{today})=1$ gives $H_0 \eta_\text{today} \approx 3.26$.   

Expanding around $\eta=0$ we find
\begin{align}
V^+ &= \frac{6}{\eta^2}+O(\eta^0), \qquad V^- = \frac{2}{\eta^2}+O(\eta^0),  
\end{align}
while around $\eta= \eta_\text{max}$, 
\begin{align}
V^+ &= \frac{3}{2}\Omega_m \sqrt{\Omega_\Lambda} H_0^3 (\eta_\text{max}-\eta)+O((\eta_\text{max}-\eta)^4), \\
V^- &= \frac{2}{(\eta_\text{max}-\eta)^2}+O((\eta_\text{max}-\eta)^1).
\end{align}
The occurrence of $1/\eta^2$ and $1/(\eta_\text{max}-\eta)^2$ fall-off conditions for the potential $V^-$ can be used to describe an approximate potential for tensor perturbations of the $\Lambda$CDM model \cite{Jokela:2022rhk}.  

The geometric optics approximation for GW propagation amounts to neglecting the potential in comparison with the wave oscillations for the tensor modes. This amounts to $\omega_\text{gw}^2 \gg V^-$. At early times $\eta \to 0$, this reduces to Eq. \eqref{GeomOpt} discussed in the Introduction. Using the early time asymptotic expansion $a=\frac{\Omega_m}{4}\eta^2+o(\eta^2)$, the geometric optics approximation equivalently amounts to 
\begin{align}
\varepsilon \equiv \frac{\frac{H_0 \Omega_m}{4}\sqrt{z_\text{emiss}}}{2\pi f_\text{gw} } \ll 1 
\end{align}
where $z_\text{emiss}=1/a(t_\text{emiss})-1$ is the redshift at emission time. This approximation is extremely well obeyed for all expected sources up to redshift $z \sim 100$ at the Einstein Telescope ($f_\text{gw} \approx 10-2000$ Hz) \cite{ET:2025xjr}  and LISA ($f_\text{gw} \approx 10^{-4}-1$ Hz) \cite{LISAConsortiumWaveformWorkingGroup:2023arg} as $\varepsilon \sim 10^{-19}$ and $\varepsilon \sim 10^{-14}$, respectively. For such sources and detectors, our relativistic formalism is unnecessary, as the geometric optics approximation can be used.

\end{enumerate}

\section{Green's function for power law cosmologies}
\label{sec:Green}

It is interesting to note that in terms of rescaled tensor, vector and scalar variables $\tilde \chi_{\langle ij \rangle }$, $\tilde \chi_{0i}$, $\tilde \chi_{ii}$ and $\tilde \chi_{00}+\tilde \chi_{ii}$ the inhomogeneous wave equations \eqref{eqspowerlaw} in power-law cosmologies have the generic form
\bea \label{1IV25.01}
\bigg(\square+\frac{C}{\eta^{2}}\bigg) \psi(x)=-4\pi \mu (x),
\eea
where $\square=\eta^{\mu \nu}\partial_{\mu} \partial_{\nu}$ is the wave operator of Minkowski space-time and $C=\alpha(\alpha \pm 1)$. We look for a retarded Green's function $G_{+}(x,x')$ such that the solution of the inhomogeneous equation \eqref{1IV25.01} can be expressed as 
\bea 
\psi(x)=\int G_{+}(x,x') \mu(x') \sqrt{-g'} d^{4} x'.
\eea
The Green's function satisfies the wave equation 
\bea \label{2IV25.04}
\bigg(\square+\frac{C}{\eta^{2}}\bigg) G_{+}(x,x')=-4\pi \delta_{4}(x,x'),
\eea
where $\delta_{4}(x,x')$ is the invariant Dirac functional 
\cite{Poisson:2011nh}. Following the Hadamard construction of the Green's 
function, we choose the ansatz
\bea  \label{04IV26.01}
G_{+}(x,x')=U(x,x') \delta_{+}(\sigma)+V(x,x') \theta_{+}(-\sigma),
\eea
where $U(x, x')$, $V(x,x')$ are smooth biscalars, and $\sigma(x,x')$ is the 
Synge world function which is half of the geodesic distance squared between the 
points $x'$ and $x$. The space-time points $x, x'$ belong to a convex normal 
neighbourhood with $x$ in the chronological future of $x{'}$. The light-cone 
delta function $\delta_{+}(\sigma)$ is supported on the future light-cone of 
$x'$, and the light-cone step function $\theta_{+}(-\sigma)$ equals to one 
inside the future light-cone. The ansatz in \eqref{04IV26.01} is generic for 
any wave operator in a curved background. Before we substitute the ansatz of 
the Green's function into the equation \eqref{1IV25.01}, we shift $\sigma$ by a 
small positive parameter $\epsilon$ so that we can perform the differentiation. 
Therefore, we consider the Green's function as 
\bea 
G_{+}^{\epsilon} (x,x')=U(x,x')\delta_{+}(\sigma+\epsilon)+V(x,x') \theta_{+}(-\sigma-
\epsilon),
\eea
and later we will recover the retarded Green's function by taking the limit $\epsilon \to 0^{+}$. Using the relation $g^{\alpha \beta}\nabla_{\alpha} \sigma\nabla_{\beta}\sigma :=g^{\alpha\beta} \sigma_{\alpha} \sigma_{\beta}=2\sigma$, and the distributional identities \cite{Poisson:2011nh}
\begin{align} \nonumber
(\sigma+\epsilon) \delta'(\sigma+\epsilon)= -\delta(\sigma+ \epsilon)-\epsilon \delta'(\sigma+\epsilon), \quad(\sigma+\epsilon) \delta''(\sigma+\epsilon)= -2\delta'(\sigma+ \epsilon)-\epsilon \delta''(\sigma + \epsilon),
\end{align}
we obtain
\begin{align} \nonumber
\bigg(\square +\frac{C}{\eta^{2}}\bigg) G_{+}^{\epsilon}&=-2\epsilon\delta_{+}''(\sigma+ \epsilon)U+2\epsilon \delta_{+}'(\sigma+ \epsilon)V+\delta_{+}'(\sigma+\epsilon) \bigg\{2U_{,\alpha} \sigma^{\alpha}+(\sigma_{\alpha}{}^{\alpha}-4)U\bigg\}\\
\nonumber
&+ \delta_{+}(\sigma+\epsilon) \bigg\{-2V_{,\alpha} \sigma^{\alpha}+(2-\sigma^{\alpha}{}_{\alpha})V+(\square+ \frac{C}{\eta^2})U\bigg\}\\ \label{2IV25.02}
&+\theta_{+}(-\sigma -\epsilon)\bigg\{(\square+\frac{C}{\eta^2})V\bigg\}.
\end{align}
In the limit $\epsilon\to 0^{+}; \epsilon \delta'(\sigma+\epsilon)\to 0$, $\epsilon \delta''(\sigma+\epsilon)\to2\pi \delta_{4}(x,x')$ \cite{Poisson:2011nh}, Eq. \eqref{2IV25.02} becomes
\begin{align} \nonumber
\bigg(\square +\frac{C}{\eta^{2}}\bigg) G_{+}&=-4\pi \delta_{4}(x,x')U+\delta_{+}'(\sigma) \bigg\{2U_{,\alpha} \sigma^{\alpha}+(\sigma_{\alpha}{}^{\alpha}-4)U\bigg\}\\ \nonumber
&+ \delta_{+}(\sigma) \bigg\{-2V_{,\alpha} \sigma^{\alpha}+(2-\sigma^{\alpha}{}_{\alpha})V+(\square+ \frac{C}{\eta^2})U\bigg\}\\ \label{2IV25.03}
&+\theta_{+}(-\sigma )\bigg\{(\square+\frac{C}{\eta^2})V\bigg\}.
\end{align}
Though in our case, $\square$ is the Minkowski wave operator, this identity also holds for wave operator in curved background.
From Eq. \eqref{2IV25.04}, the right-hand side of Eq. \eqref{2IV25.02} should be $-4\pi \delta_{4}(x,x')$. This gives the coincidence limit, as $x\to x'$
\bea \label{2IV25.05}
\lim_{x \to x'}U(x,x')=:[U]=1
\eea
for the biscalar $U(x,x')$. We will consider the elimination of all terms sequentially. To eliminate the $\delta_{+}'$ term, we have
\bea \label{2IV25.06}
2U_{,\alpha} \sigma^{\alpha}+(\sigma_{\alpha}{}^{\alpha}-4)U=0. 
\eea
These two equations determine $U(x,x')$ uniquely as \cite{Poisson:2011nh}
\bea 
U(x,x') =\Delta^{1/2}(x,x'),
\eea
where $\Delta$ is van Vleck determinant defined as 
\bea 
\Delta (x,x')=-\frac{\mbox{det}[-\sigma_{\alpha'\beta}(x,x')]}{\sqrt{-g}\sqrt{-g'}},
\eea
where $g$, and $g'$ are the metric determinant at $x$ and $x'$ respectively. 
For our case, the Synge's world function is $\sigma(x,x')=\frac{1}{2}
\eta_{\alpha \beta}(x-x')^{\alpha}(x-x')^{\beta}$. Hence in the coincident 
limit, $[\sigma_{\alpha'\beta}]=\eta_{\alpha' \beta'}$. Therefore, for the wave 
operator in Minkowski space-time $U(x,x')=1$.

The $\delta_{+}$ term in Eq. \eqref{2IV25.03} can be eliminated by demanding 
that its coefficient vanish when $\sigma=0$. Denoting $\check{V}(x,x')$ as the 
restriction of $V(x,x')$ on the light cone, we have
\bea \label{15IV25.01}
-2\check{V}_{,\alpha} \sigma^{\alpha}+(2-\sigma^{\alpha}{}_{\alpha})\check{V}=-
\frac{C}{\eta^2}.
\eea
This equation can be rewritten as 
\bea \label{9IX25.01}
\lambda \frac{d \check{V}}{d\lambda}+\check{V}-\frac{C}{2\eta^{2}}=0,
\eea
where $\lambda$ is the affine parameter along the null geodesic. We have also 
used $\sigma^{\alpha}{}_{\alpha}=4$ for the Minkowski metric. One can choose 
$\lambda:=\eta-\eta'$ as affine parameter along the null cone. Therefore, Eq. 
\eqref{15IV25.01} becomes
\bea \label{23IV25.01}
(\eta-\eta') \frac{d \check{V}}{d\eta}+\check{V}-\frac{C}{2\eta^{2}}=0. 
\eea
The solution to Eq. \eqref{23IV25.01} is
\bea \label{9IX25.02}
\lambda \check {V}=-\frac{C}{2} \frac{1}{\lambda+ \eta'}+K (x'),
\eea
where $K(x')$ is the integration constant. This integration constant can be obtained from the coincidence limit of Eq. 
 \eqref{15IV25.01}. In the coincidence limit $x \to x'$, $[\sigma^{\alpha}]=0$. Therefore, the coincidence limit of $\check V$ is 
\bea \label{22V25.01}
\lim_{x \to x'} \check {V}(x,x')=:[V]=\frac{C}{2\eta'^{2}}.
\eea
$\check{V}(x,x')$ is regular in the coincidence limit, and the coincidence limit of $[\lambda]=0$. Therefore, the integration constant in equation \eqref{9IX25.02} can be given as 
\bea 
K(x')=\frac{C}{2\eta'}.
\eea
Therefore, the solution to Eq. \eqref{23IV25.01} becomes
\bea \label{8IX25.03}
\check{V}(x,x')=\frac{C}{2}\frac{1}{\eta \eta'}.
\eea
In summary, Eq. \eqref{9IX25.01} and the initial condition at the tip of the light-cone \eqref{22V25.01} have uniquely fixed $\check{V}(x,x')$ along the light-cone.

The elimination of the $\theta_{+}$ term in Eq. \eqref{2IV25.03} gives 
\bea  \label{17IV25.01}
(\square+\frac{C}{\eta^2})V(x,x')=0.
\eea
We have to look for a solution of the characteristic initial value problem that satisfies both Eqs. \eqref{23IV25.01} and \eqref{17IV25.01} simultaneously. This problem can be solved by a series \cite{Friedlander:2010eqa}\footnote{We thank Abraham Harte for pointing out this reference.}
\bea \label{25IX25.01}
V=\sum_{n = 0}^{\infty} V_{n} \frac{\sigma^{n}}{n !},
\eea
where $V_{0}$ is identified with $\check{V}$ on the light-cone and where $V_{n}$, for $n\geq 1$, satisfy a set of transport equations. Using $\sigma_{\alpha} \sigma^{\alpha}=2\sigma$, and denoting $P \equiv \square +\frac{C}{\eta^{2}}$, we note the identity
\bea \label{8IX25.01}
P\bigg(V_{n} \frac{\sigma^{n}}{n !}\bigg)=P(V_{n}) \frac{\sigma^{n}}{n !}
+\bigg(2(\partial^{\alpha}V_{n})(\partial_{\alpha} \sigma)+(\sigma^{\alpha}{}_{\alpha}+2n -2)V_{n}\bigg)\frac{\sigma^{n-1}}{(n -1)!}.
\eea
Suppose that $V_{n}$ can be chosen so that 
\bea \label{8IX25.02}
2(\partial^{\alpha}V_{n})(\partial_{\alpha} \sigma)+(\sigma^{\alpha}{}_{\alpha}+2n -2)V_{n}=-P(V_{n -1}).
\eea
Therefore, Eq. \eqref{8IX25.01} becomes
\bea 
P\bigg(V_{n}\frac{\sigma^{n}}{n !}\bigg)=P(V_{n})\frac{\sigma^{n}}{n !}-P(V_{n-1})\frac{\sigma^{n -1}}{(n -1)!}.
\eea
The recurrence relations in Eq. \eqref{8IX25.02} determine a sequence of functions $V_{n}$. Assuming that $V_{n-1}$ has already been determined, $V_{n}$ can be obtained from the transport equation \eqref{8IX25.02}.  By consistency of the coincidence limit, $x \to x'$, we impose that $V_{n}$ is bounded in that limit as a boundary condition.  To find a generic solution for $V_{n}$, it is useful to rewrite \eqref{8IX25.02} as
\bea 
\lambda \frac{dV_{n}}{d\lambda}+ (n+1) V_{n}=-\frac{1}{2}P(V_{n-1}),
\eea
where we have used $\sigma^{\alpha}{}_{\alpha}=4$. After multiplying with the integrating factor $\lambda^{n}$, this equation can be written as
\bea 
\frac{d}{d\lambda}(\lambda^{n+1} V_{n})=- \frac{1}{2} \lambda^{n} P(V_{n-1}).
\eea
Therefore, the solution for $V_{n}$ is
\bea \label{10IX25.01}
V_{n}=-\frac{1}{2} \frac{1}{\lambda^{n+1}} \int_{0}^{\lambda} d\lambda \lambda^{n} P(V_{n-1}).
\eea
We note that with the identification 
$V_{-1}=U$ and the already defined $V_{0}=\check{V}$, Eq. \eqref{10IX25.01} for $n=0$
reproduces Eq. \eqref{8IX25.03} along the light-cone. The next order solution $n=1$ can be obtained as follows. 
We note that 
\bea 
P(V_{0}) &\equiv& \bigg(\square + \frac{C}{\eta^{2}}\bigg) V_{0}
\label{10IX25.02}
= \frac{C (C-2) }{2} \frac{1}{\eta' \eta^{3}}.
\eea
After a straightforward algebra with the substitution $\lambda= \eta-\eta'$, the solution for $V_{1}$ can be obtained as
\bea
V_{1}=- \frac{C(C-2)}{8} \frac{1}{(\eta' \eta)^{2}}.
\eea  
From Eq. \eqref{10IX25.01}, a generic solution for $V_{n}$, $n \geq 0$, can be written as
\bea
V_n =- \left(-\frac{1}{2\eta' \eta} \right)^{n+1}\prod_{i=0}^n \frac{C-i(i+1)}{i+1},
\eea
which is bounded in the coincidence limit.  In the case where $C$ takes the 
form $C=l (l+1)$ for $l$ a positive integer, $V_m=0$ for all $m \geq l$ so the 
solution truncates when $l$ is a positive integer. Also, when  $l$ is a 
negative integer, $V_m=0$ for all $m \geq -l-1$ so the solution also truncates 
when $l$ is a negative integer.

To understand this series expansion in an analytic form for a generic $C$, it is judicious to introduce the ansatz for $V(x,x')$ as
\bea \label{17IX25.01}
V(x,x')= \frac{1}{\eta \eta'} F(p), \quad \mbox{where} \quad p=\frac{\sigma}{\eta \eta'}.
\eea
Note that this ansatz satisfies the reciprocity property of the Green's function $V(x,x')=V(x',x)$. 
With this ansatz, the equation for V
\bea
\bigg(\square+ \frac{C}{\eta^{2}}\bigg) V(x,x')=\bigg(-\partial_{\eta}^{2} +\frac{2}{R} \partial_{R}+\partial_{R}^{2}+ \frac{C}{\eta^{2}}\bigg)V(x,x')=0,
\eea
where $R= \vert x-x' \vert$, becomes
\bea \label{17IX25.02}
\frac{1}{\eta^3 \eta'}\bigg(p(2-p) \frac{\partial^{2}F}{\partial^{2}p}+ 4(1-p) \frac{\partial F}{\partial p}+(C-2)F\bigg)=0. 
\eea
Note that on the light-cone, $p=0$. Therefore, from \eqref{8IX25.03}, and \eqref{17IX25.01}, we have 
\bea 
F(p)|_{p=0}=\frac{C}{2}.
\eea
With this initial condition on the light-cone, one discards the irregular Meijer 
solution to \eqref{17IX25.02} and the solution of \eqref{17IX25.02} is obtained 
as the hypergeometric function 
\bea \label{hyper}
F(p)= \frac{C}{2}  \ _{2}F_{1}\bigg(\frac{3-\sqrt{1+4C}}{2},\frac{3+\sqrt{1+4C}}{2}; 2; \frac{p}{2}\bigg).
\eea
We checked that the power series expansion  of $V$ in Eq. \eqref{17IX25.01} 
matches with the light-cone expansion of $V$ in Eq. \eqref{25IX25.01} order by 
order. Hence, the Green's function of the differential operator in equation \eqref{1IV25.01} is given by
\bea \label{26IX25.02}
G_{+}(x,x')= \delta_{+}(\sigma)+\frac{C}{2 \eta \eta'}  \ _{2}F_{1}\bigg(\frac{3-\sqrt{1+4C}}{2},\frac{3+\sqrt{1+4C}}{2}; 2; \frac{\sigma}{2\eta \eta'}\bigg) \theta_{+}(-\sigma).
\eea
After using the property $ \ _{2}F_{1}(x,y;z;t)= \ _{2}F_{1}(y,x;z;t)$ and substituting $C= \alpha(\alpha \pm 1)$, the final Green's function solution to Eq. \eqref{2IV25.04} is 
\begin{align}\label{OurGreen}
G_{+}(x,x')= \delta_{+}(\sigma)+\frac{\alpha(\alpha \pm 1)}{2 \eta \eta'}  \ _{2}F_{1}\big(2 \pm \alpha , 1 \mp \alpha ; 2; \frac{\sigma}{2\eta \eta'}\big) \theta_{+}(-\sigma) . 
\end{align}

For our purposes, it is also useful to write down the Green's function in the following form 
\bea \label{26IX25.03}
G_{+}(x,x')= \delta_{+}(\sigma)+\bigg(\frac{\alpha(\alpha \pm 1)}{2 \eta \eta'} +\sum_{n=1}^{\infty} V_{n} \frac{\sigma^n}{n!}  \bigg)\theta_{+}(-\sigma). 
\eea
On the one hand, the $\delta_{+}$ part of 
the Green's function is exactly same in a generic spatially 
flat FLRW model simply because the gravitational perturbations obey wave equations with a Minkowskian D'Alembertian. The Synge's world 
function $\sigma(x,x')$ is universal because it relies on the Minkowski background structure. On the other hand, the $\theta_{+}$ part or tail part is specific to power-law cosmologies since we made use of the $1/\eta^2$ scaling of the potential. For a generic FLRW model, one would need to generalize our analysis. 

Let us now compare our result with the literature in chronological order. 

The Green's function of matter dominated universe was studied by Waylen \cite{Waylen:1978dx, a2540274-da7b-3e43-b5e5-9617401299cd} using the Hadamard construction. However, his analysis used a formulation of linearized Einstein's equations which is valid for a maximally symmetric background but which is incorrect for an FRLW background, see his Eq. (4) of \cite{Waylen:1978dx}. Later on, the Green's functions were studied in \cite{Caldwell:1993xw} for a generic FLRW universe. However, even for  transverse, traceless, synchronous perturbations (which actually cannot be obtained as gauge conditions in the presence of sources), our perturbed equation in Eq. \eqref{6II25.01} does not match with that of Eq. (3.3) in \cite{Caldwell:1993xw}. Our Green's functions therefore disagree with these results.

The Green's function of Eq. \eqref{1IV25.01} has been obtained in \cite{Haas:2004kw} in the context of the propagation of a scalar 
charge in power-law cosmologies. In \cite{Haas:2004kw}, a generic solution in the case $C=l(l+1)$ with $l$ integer for 
$V(x,x')$ was obtained in terms of the Appell $F_{4}$ function of two variables, 
\bea \label{26IX25.01}
V(x,x')= \frac{(2l-1)!!}{(2l-2)!!} \frac{(\eta/\eta')^{l}}{\eta^2} F_{4}(-l+1, 
\frac{3}{2}; -l+\frac{1}{2}, \frac{3}{2}; (\eta'/\eta)^{2}, (R/\eta)^{2}).
\eea
We checked that for any positive or negative integer values of $l$, our solution of $V(x,x')$ in Eqs. \eqref{17IX25.01}-\eqref{hyper} matches with that of Eq. \eqref{26IX25.01}. Our solution is more simply expressed as a hypergeometric function with variable $\sigma/\eta\eta'$ instead of a Appell function of two variables. Moreover, the solution \eqref{26IX25.01} is singular on the light-cone for complex $l$. Our solution \eqref{17IX25.01}-\eqref{hyper} is instead regular for arbitrary $C$. 

The Green's function was constructed by Chu in Appendix B of \cite{Chu:2016qxp} and derived for power law cosmologies in Eq. (116) of \cite{Chu:2016ngc}. For the wave operator with $C=\alpha(\alpha \pm 1)$, we find that we can rewrite the Green's function \eqref{OurGreen} as 
\begin{align}
G_+(x,x') = \delta_+(\sigma) - \frac{d}{d\sigma}P_{\pm \alpha}[1-\frac{\sigma}{\eta \eta'}]\theta_+(-\sigma),   
\end{align}
where $P_{\pm \alpha}[x]$ are the Legendre polynomials. This exactly matches with \cite{Chu:2016qxp,Chu:2016ngc} after switching to mostly minus convention. We therefore confirm the Green's functions of Chu \cite{Chu:2016qxp,Chu:2016ngc} with our independent derivation.

For completeness, we note that the Green's function was also computed in the Fourier domain in \cite{Faraoni:1991xe,Iliopoulos:1998wq} in terms of Hankel functions. Since our derivation is in position space, we cannot directly compare our expressions with their results.

\section{Multipolar decomposition of the stress-energy tensor}
\label{sec:multi}

We aim to study compact sources which are characterized by their most relevant multipolar 
components. Before solving the inhomogeneous wave equations, we will start by 
generalizing the multipolar decomposition of the stress-energy tensor around a de 
Sitter background \cite{Compere:2023ktn} to the case of a generic spatially flat FLRW 
background. We will then be able to perform a quadrupolar truncation. 


Integrals of local densities are best defined using tensors expressed in an orthonormal frame tetrad. The frame components are coordinate scalars but transform under the Lorentz group acting on the frame. For the homogeneous metric \eqref{21IX25.01}, we define the comoving coordinates $\bar x_\alpha=a(\eta)x_\alpha$, the comoving tetrad frame $f^{\alpha}_{\bar \alpha} = a(\eta)^{-1} \delta_{\bar \alpha}^{\alpha}$ defined such as $\eta_{\bar \alpha \bar \beta}= f^{\alpha}_{\bar \alpha}f^{\beta}_{\bar \beta}\bar g_{\alpha\beta}$ and its inverse $f_{ \alpha}^{\bar \alpha} = a(\eta) \delta_{ \alpha}^{\bar \alpha}$. 

Given the role of the shifted stress-energy tensor defined in \eqref{tildeT}, we define the moments of the energy density $Q_L^{(\rho)}$, of the momentum density $P_{i \vert L}$ and of the stress density $S_{ij \vert L}$ as  
\begin{align} 
Q_L^{(\rho)}(\eta)&:=\int d^3 \bar x \delta \tilde T_{\bar 0 \bar 0}\bar x_L = \int d^3x a^{\ell + 1}(\eta) \delta \tilde T_{00} x_L,\label{25IV23.02} \\ 
P_{i \vert L}(\eta)&:=\int d^3 \bar x \delta \tilde T_{\bar 0 \bar i}\bar x_L = \int d^3x a^{\ell + 1}(\eta) \delta \tilde T_{0i} x_L, \label{25IV23.02bis}\\
S_{ij \vert L}(\eta) &:=\int d^3 \bar x \delta \tilde T_{\bar i \bar j} \bar x_L = \int d^3x a^{\ell + 1}(\eta) \delta \tilde  T_{ij}  x_L.\label{27IV23.01}
\end{align} 
We denote $L=i_{1}i_{2}\cdots i_{\ell}$ as a multi-index made of $\ell$ spatial indices, and $x_{L}=x_{i_{1}}\cdots x_{i_{\ell}}$. Such set of moments completely characterize the variations of the stress-energy tensor.  For convenience, we also introduce the moments of the pressure density $Q_L^{(p)}$ as
\begin{align} 
\label{25IV23.03}
Q_L^{(p)}(\eta) &=\int d^3 \bar x \eta^{\bar i \bar j} \delta \tilde T_{\bar i \bar j} \bar x_L = \int d^3x a^{\ell + 1}(\eta) \delta_{ij}\delta \tilde T_{ij} x_L=S_{ii \vert L} ,
\end{align}
and we further define 
\begin{align} 
\label{25IV23.03bis}
Q_L^{(\rho+p)}(\eta) :=Q_L^{(\rho)}(\eta)+Q_L^{(p)}(\eta). 
\end{align}

\subsection{Conservation equations}

The conservation of the stress-energy tensor on the homogeneous spacetime \eqref{21IX25.01} is equivalent to the flux-balance laws \eqref{tildecons}. Let us rewrite it here as 
\begin{subequations}\label{tildecons2}
\begin{align}
\partial_\eta(a\delta  \tilde T_{00}) - \partial_i (a \delta \tilde T_{0i})+\dot a \delta \tilde T_{kk} &=s(\eta) a^2 \hat \chi,\label{cons00} \\ 
(\partial_\eta+\frac{\dot a}{a})(a \delta \tilde T_{0i})-\partial_j (a \delta \tilde T_{i j}) &=0.   
\end{align}    
\end{subequations}
where the background cosmological source $s(\eta)$ is defined as $s(\eta)\equiv  -\frac{1}{4a^2}\partial_\eta( a^2(p+\epsilon) )$ and we defined $\hat \chi \equiv \tilde \chi_{00}+\tilde \chi_{kk} $. Let us convert $\partial_{\eta}$ to $\partial_{t}$ using the relation $\partial_{\eta}=a(t) \partial_{t}$ while remaining in the conformal coordinates $(\eta,\vec{x})$.  This leads to 
\begin{align}   \label{28II23.01}
 a \partial_t \delta\tilde T_{00}(\eta,x)-\partial_i \delta\tilde  T_{0i}(\eta,x)+(\partial_{t}a)( \delta\tilde T_{00}(\eta,x)+\delta_{ij} \delta\tilde T_{ij}(\eta,x)) &=s(\eta) a \hat \chi , \\ \label{28II23.02}
a\partial_t  \delta\tilde T_{0i}(\eta,x)-\partial_j  \delta\tilde T_{ij}(\eta,x)+2({\partial_t}a)  \delta\tilde T_{0i}(\eta,x)&=0.
 \end{align}
Note that the $0$ index means the $\eta$ component. Multiplying these equations by $a^\ell(\eta)x_L$, integrating over $\vec{x}$ and using the definition of moments of the stress-energy tensor we have
\begin{align}
    \partial_t Q_L^{(\rho)} &= H(\ell Q_L^{(\rho)}-Q_L^{(p)})- \ell  P_{(i_1|i_2 \cdots i_\ell)} +s(\eta) \hat \chi_L\ ,\label{eq37}\\ 
    \partial_t P_{i|L} &= (\ell-1) H P_{i|L} - \ell  S_{i(i_1|i_2 \cdots i_\ell)},\label{eq38}
\end{align}
where $H:=\frac{\partial_{t}a}{a}$ is the Hubble parameter, $S_{i\emptyset}:=0$ by definition and the moments of the scalar cosmological source are defined as 
\begin{align}
   \hat \chi_L := \int d^3x a^{\ell +1}(\eta) \hat \chi x_L. 
\end{align}

These equations have the same form as previously derived around de Sitter, see Eqs. (2.37)-(2.38) of \cite{Compere:2023ktn} except that in FRLW universe, one has generically an additional source term proportional to $s(\eta)$. We define the angular momentum (or odd parity dipole moment) as 
\begin{align}
J_i = \epsilon_{ijk}P_{j \vert k},
\end{align}
which is conserved, $\partial_ t J_i =0$, in the linear theory as a consequence of the $SO(3)$ invariance of the FRLW background.  Using Eqs \eqref{eq37} and \eqref{eq38} one obtains
\begin{align} 
S_{ij}(\eta)  =\int d^{3}x' a(\eta)\delta \tilde{T}_{ij} (\eta, x') = \frac{1}{2}\partial_t \left( \partial_t Q_{ij}^{(\rho)} - 2H Q_{ij}^{(\rho)}+H Q_{ij}^{(p)}-s \hat \chi_{ij}\right).\label{Sijpartial} 
\end{align}
Solving Eq. \eqref{eq38} with $\ell \mapsto \ell+1$ in terms of $S_{i(j \vert L)}$ we obtain
\begin{align}
S_{i (j \vert L )} &= -\frac{1}{\ell +1}(\partial_t - \ell H)P_{i \vert j L}.\label{eqn37}
\end{align}
The totally symmetric part $S_{(ij \vert L)}$ is a function of $P_{(i\vert j L)}$ which can be substituted using Eq. \eqref{eq37} with $\ell \mapsto \ell + 2$. The two conservation equations \eqref{eq37}-\eqref{eq38} are therefore equivalent to Eq. \eqref{eqn37} and 
\begin{align}\label{SijL}
S_{(ij \vert L)} &= \frac{1}{(\ell+1)(\ell+2)}(\partial_t - \ell  H)\left( (\partial_t - (\ell+2)H)Q^{(\rho)}_{ij L}+H Q^{(p)}_{ij L}-s \hat \chi_{ijL}\right),
\end{align}
where Eq. \eqref{Sijpartial} is recovered when $\ell=0$, in addition to the cases $\ell=0,1$ of Eq. \eqref{eq37} and $\ell=0$ of Eq. \eqref{eq38} which cannot be inverted for $S_{ij \vert L}$. These lower order equations provide the non-conservation of the 7 quantities associated with broken symmetries of the FRLW background, 
\begin{align}
 \partial_t Q^{(\rho)} &= - H Q^{(p)}+s \hat \chi_\emptyset, \\ 
 \partial_t Q^{(\rho)}_i &= H(Q^{(\rho)}_i - Q^{(p)}_i)-P_i +s \hat \chi_i ,\\ 
 \partial_t P_i &= - H P_i . 
\end{align}

Note finally that contracting the Eq. \eqref{SijL} with $\delta_{ij}$ we obtain
\begin{align}
    2 Q_L^{(p)} &=(\partial_t -\ell H) (\partial_t Q_{kkL}^{(\rho)}+ H (Q_{kkL}^{(p)}-(\ell+2)Q_{kkL}^{(\rho)}) -s \hat \chi_{ii L})  \nonumber\\ 
   & - \ell (\ell-1) S_{(i_1i_2|...i_\ell)kk} -4\ell S_{k(i_{1}|\cdots i_{\ell})k} ,\label{CONSCOND}
\end{align}
which generalizes Eq. (2.50) of \cite{Compere:2023ktn} obtained for a de Sitter background to a generic spatially flat FRLW background.

\subsection{Consistent quadrupolar truncation} 

In all subsequent sections, we will restrict our analysis to the quadrupolar order, which captures the dominant effects of gravitational radiation. The quadrupolar truncation was consistently defined in \cite{Compere:2023ktn}. It amounts to fixing 
\begin{equation}
\int d^3 x a^{\ell+1} T_{\mu\nu}x^L=0,\qquad \forall \ell >2. 
\end{equation}
The equation of motion \eqref{18IX25.01tilde} then implies that $\chi_L=0$ for all $\ell >2$. All definitions provided in \cite{Compere:2023ktn} in the case of de Sitter background then extend straightfowardly to the case of a background FRLW spacetime. 

Explicitly, in the quadrupolar truncation, the conservation equations \eqref{eq37}-\eqref{eq38} imply
\begin{align}
P_{(i \vert j k )}=0, \qquad P_{i \vert j k l}=0,\qquad S_{i (j \vert kl )}= 0.
\end{align}
Eq. \eqref{SijL} then imply $S_{(ij \vert k)}=0$. 
These conditions imply the trace conditions
\begin{equation}
P_{j\vert ji}=-\frac{1}{2}P_{i\vert jj}, \quad 
S_{ij \vert j}=-\frac{1}{2}S_{jj\vert i}=-\frac{1}{2}Q^{(p)}_{i}, \quad 
S_{ij \vert kk}=S_{kk \vert ij}=Q^{(p)}_{ij},\quad S_{i k \vert k j}= - \frac{1}{2}Q^{(p)}_{ij}. 
\end{equation}
The tensors $\epsilon_{ikl} S_{jk \vert l}$ and $\epsilon_{ilm}S_{j l\vert m k}$ 
are traceless with respect to each index contraction. Their antisymmetric parts are determined in terms of $Q^{(p)}_i$ and $Q^{(p)}_{ij}$. The symmetric tracefree part of $\epsilon_{ikl} S_{jk \vert l}$ is unconstrained. Now, the symmetric tracefree part of $\epsilon_{ilm}S_{j l\vert m k}$ reads explicitly as
\begin{equation}
\epsilon_{ilm} (S_{j l\vert m k}+S_{k l\vert m j})+ \epsilon_{jlm}(S_{k l\vert m i} +S_{i l\vert m k})+\epsilon_{klm} (S_{i l\vert m j}+ S_{j l\vert m i})=0, \label{eq:antisym S}
\end{equation}
which identically vanish after using $S_{l(i|jm)}=0$ and the fact that $S_{ij \vert kl}$ is symmetric in both $ij$ and $kl$ indices. We can then perform the following decompositions in terms of irreducible tensors  
\begin{subequations}\label{decom}
    \begin{align}
P_{i \vert jk} &=\frac{1}{2}\epsilon_{li(j}J_{k)l}-\frac{1}{2}\delta_{i(k}P_{j) \vert ll}+\frac{1}{2}\delta_{jk}P_{i \vert ll},\\
S_{i j \vert k} &= \frac{1}{2}\epsilon_{kl (i} K_{j)l}-\frac{1}{2}\delta_{k(i}Q^{(p)}_{j)}+\frac{1}{2}\delta_{ij}Q^{(p)}_k, \\
S_{i j \vert kl} &= \delta_{ij}Q^{(p)}_{kl}- (\delta_{i(k}Q^{(p)}_{l)j}+\delta_{j(k}Q^{(p)}_{l)i})+ Q^{(p)}_{ij}\delta_{kl}-\frac{1}{2} \delta_{ij}\delta_{kl}Q^{(p)}_{mm}+\frac{1}{2} \delta_{i(k}\delta_{l)j}Q^{(p)}_{mm},\label{Sijkl}
\end{align}
\end{subequations}
where we defined the symmetric and tracefree odd parity quadrupolar moments 
\begin{align}
J_{ij} &:=\frac{4}{3}P_{k \vert l (i}\epsilon_{j) kl}, \\
K_{ij}& :=\frac{4}{3}\epsilon_{kl(i} S_{j)k \vert l}. 
\end{align}
Taking the trace and the antisymmetric part in $ij$ of Eq. \eqref{eq38} for $L=jk$, we obtain
\begin{align}\label{consJK}
(\partial_t - H)P_{i \vert jj} &=Q_i^{(p)},\\ 
(\partial_t - H)J_{ij} &= - K_{ij}.
\end{align}

\section{Quadrupolar perturbations in power law cosmologies}
\label{sec:quad}

A generic scalar, vector and tensor perturbation of a power-law cosmology obeys Eq. \eqref{1IV25.01} which we repeat here for convenience 
\bea \label{9XI25.01}
\bigg(\square + \frac{\alpha (\alpha \pm 1)}{\eta^{2}}\bigg)\psi(x)=-4\pi \mu(x) .
\eea 
Using the retarded Green's function derived in Section \ref{sec:Green} and using the identity 
\bea 
\delta_{+}(\sigma)=\frac{\delta (\eta'-\eta +|\bar{x}-\bar{x}'|)}{(d\sigma/d\eta')|_{{\eta'}=\eta -|\bar{x}-\bar{x}'|}  }=\frac{\delta (\eta'-\eta +|\bar{x}-\bar{x}'|)}{|\bar{x}'-\bar{x}|},
\eea
the inhomogeneous solution can be written as
\begin{align}  \label{29IX25.01}
\psi(x)= \int d^{3}x' \frac{\mu(\eta - |\bar{x}-\bar{x}'|, \bar{x}')}{|\bar{x}-\bar{x}'|}
+  \sum_{n=0}^{\infty}\int d^{3} x' \int_{\eta_{i}}^{\eta - |\bar{x}-\bar{x}'|} d\eta' \frac{V_{n} \sigma^{n}}{n!}  \mu(\eta', \bar{x}').
\end{align}
We will call the first term the lightcone integral and the second term the tail integral. The existence of a tail integral is a generic property of  
wave propagation in a curved background. Due to the background curvature, the waves backscatter and the propagation can 
be inside the light-cone. In our case, the 
tail term is generated already at linear order in the perturbative expansion around the FLRW background. This feature is reflected in the Heaviside function in 
the Hadamard Green's function. As a consequence, the field at a particular spacetime 
point does not only depend upon the response of the source at the corresponding
retarded instant of time. Rather, it depends upon the entire past history of the source. Usually, it is challenging to obtain a closed form expression of the tail 
integral in a generic context. 

\subsection{Tail integral up to quadrupolar order}

In this section, we will provide an analytic expression of the 
tail integral in Eq. \eqref{29IX25.01} for power-law cosmologies up to quadrupolar order.

For the purpose of this section, we define the moments of any function $\mu(\eta, {x})$ in FLRW geometry as
\bea  \label{defQL}
Q_{L}(\eta):= \int d^{3} x a^{\ell }(\eta) \mu(\eta, x) x_{L}.
\eea
The scalar, vector and tensor perturbations correspond to either $\mu=4G a (\delta\tilde T_{00}+\delta \tilde T_{ii})$ or $\mu=4G a \delta \tilde T_{ii}$ , $\mu=4G a \delta \tilde T_{0i}$ and $\mu=4G a \delta \tilde  T_{\langle ij \rangle }$, respectively, and $Q_L(\eta)$ then reduce to the respective multipolar moments defined in Section \ref{sec:multi}.

We denote as $d$ the spatial dimension of the localized source and as  $\rho$ the comoving distance of the source with respect to the observer. We consider that the source has a dimension much smaller than its cosmological distance to us, $d/\rho \ll 1$. 
We start by rewriting the tail integral using 
\begin{align}\label{int2}
 \int_{\eta_i}^{\eta - |\bar{x}-\bar{x}'|} f(\eta')d\eta'= \int_{\eta_i}^{\eta- \rho} f(\eta')d\eta' + \int_0^{\rho - |\bar{x}-\bar{x}'|} f(\eta'+\eta_{\text{ret}})d\eta', 
\end{align}
where the second $\eta'$ has been shifted by $\eta_\text{ret} := \eta-\rho$. After splitting the integral, we will denote the resulting tail terms as $T_1$ and $T_2$, respectively. 

In the quadrupolar approximation, we limit our analysis to the terms up to quadrupolar order in $d/\rho$, 
\begin{align}\label{xxp}
|\vec{x}-\vec{x}'|
&= \rho\bigg(1 - \frac{\vec{n} \cdot \vec{x}'}{\rho} + \frac{r'^2-(\vec{n} \cdot \vec{x}')^2}{2\rho^2}+O(\frac{d^3}{\rho^3})\bigg),\\ \label{17X25.05}
|\vec{x}-\vec{x}'|^{-1}
&=\frac{1}{\rho}\bigg(1 +\frac{\vec{n} \cdot \vec{x}'}{\rho} - \frac{r'^2-3(\vec{n} \cdot \vec{x}')^2}{2\rho^2}+O(\frac{d^3}{\rho^3})\bigg). 
\end{align}
We now perform the quadrupolar expansion of  $\sigma{(\eta', x')}$,
\bea 
\sigma &=& \sigma_{\rho} -\rho \vec{n} \cdot \vec{x}'+\frac{1}{2} r'^{2} + \cdots, \label{19X25.02}
\eea
where we define $\sigma_\rho := \frac{1}{2}(-(\eta-\eta')^{2}+ \rho^{2})$. It implies that for any positive integer $n$
\bea
\sigma^{n}&=& \sigma_{\rho}^{n}-n\rho \sigma_{\rho}^{n-1}\vec{n}\cdot \vec{x}'+\frac{1}{2}(n\sigma_{\rho}^{n-1}r'^{2}+n(n-1)\sigma_{\rho}^{2}\rho^{2}(\vec{n}\cdot \vec{x}')^{2})+ \cdots.
\eea
Plunging these expansions into the tail term of Eq. \eqref{29IX25.01}, using the definition of multipolar moments \eqref{defQL} and using the resummation 
\bea 
\sum_{n=0}^{\infty} \frac{1}{n!} V_{n} \sigma_{\rho}^{n}=\frac{\alpha(\alpha \pm 1)}{2 \eta \eta'}  \ _{2}F_{1}\big(2 \pm \alpha , 1 \mp \alpha ; 2; \frac{p_\rho}{2}\big), \  \ \mbox{with } \ p_{\rho}:= \frac{\sigma_{\rho}}{\eta \eta'},
\eea
we find 
\begin{align}
T_1 = &\sum_{n=0}^{\infty} \int d^{3}x'\int_{\eta_{i}}^{\eta-\rho} d\eta' \frac{V_{n}(\eta')}{n!} \sigma^{n}(\eta', \bar{x}') \mu(\eta', \bar{x}'), \label{514}\\ \nonumber
&=\int_{\eta_{i}}^{\eta -\rho} d\eta' \frac{\alpha(\alpha \pm 1)}{2\eta \eta'}\bigg[\ _{2}F_{1}\big(2 \pm \alpha , 1 \mp \alpha ; 2; \frac{p_\rho}{2}\big) Q(\eta')\\ \nonumber
&-\frac{1}{a(\eta')} \frac{\rho}{\eta \eta'} \frac{d}{dp_{\rho}} \bigg( \ _{2}F_{1}\big(2 \pm \alpha , 1 \mp \alpha ; 2; \frac{p_\rho}{2}\big)\bigg) n_{k} Q_{k}(\eta')\\ \nonumber
&+ \frac{1}{a^{2}(\eta')} \frac{1}{2} \frac{1}{\eta \eta'}  \frac{d}{dp_{\rho}} \bigg(\ _{2}F_{1}\big(2 \pm \alpha , 1 \mp \alpha ; 2; \frac{p_\rho}{2}\big)\bigg) \delta_{kl} Q_{kl}(\eta')\\ 
&+\frac{1}{a^{2}(\eta')} \frac{1}{2} \frac{\rho^2}{(\eta \eta')^{2}}  \frac{d^{2}}{d^{2}p_{\rho}} \bigg( \ _{2}F_{1}\big(2 \pm \alpha , 1 \mp \alpha ; 2; \frac{p_\rho}{2}\big)\bigg) n_{k} n_{l} Q_{kl}(\eta')
\bigg].
\end{align}
Now, let us look into the second term $T_2$ of the split of the integral \eqref{int2}. At quadrupolar order, we can expand 
\bea 
\delta := \rho - |\bar{x}-\bar{x}'|= \vec{n} \cdot \vec{x}' -\frac{r'^{2}-(\vec{n}\cdot \vec{x}')^{2}}{2\rho}+ \cdots.
\eea
and neglect higher order terms. The non-vanishing terms at quadrupolar order of the second part according to the split \eqref{int2} of the tail term in Eq. \eqref{29IX25.01} are  
\begin{align} 
T_2 =  &\sum_{n=0}^{\infty}\int d^{3} x'\int_{0}^{\rho-|\bar{x}-\bar{x}'|} d\eta' \frac{V_{n}(\eta_{\text{ret}}+\eta')}{n!} \sigma^{n}(\eta_{\text{ret}}+\eta', \bar{x}') \mu(\eta_{\text{ret}}+\eta', \bar{x}'),\\ \nonumber
 & = \sum_{n=0}^{\infty}\int d^{3} x'\int_{0}^{\rho-|\bar{x}-\bar{x}'|} d \eta'\bigg[\{V_{n}(\eta_{\text{ret}})+\eta'V_{n}^{(1)}+\cdots\} \{\sigma^{n}(\eta_{\text{ret}})+ \eta' (\sigma^{n})^{(1)}+ \cdots\}\\ \label{19X25.01}
& \hspace{8.0cm}\times\{\mu(\eta_{\text{ret}})+ \eta' \mu^{(1)}+ \cdots\}
 \bigg].
\end{align}
More precisely, only the terms at most linear in $\eta'$ in the Taylor expansion of the integrand \eqref{19X25.01} around $\eta'=0$ will contribute at quadrupolar order. From Eq. \eqref{19X25.02}, we also note that 
\bea 
\sigma|_{\eta_{\text{ret}}}&=&-\rho \vec{n}\cdot \vec{x}'+ \frac{1}{2} r'^{2}+\cdots,\\
(\sigma)^{(1)}|_{\eta_{\text{ret}}}&=& \rho+ \cdots,\\
\sigma^{2}|_{\eta_{\text{ret}}}&=& \rho^{2}(\vec{n}\cdot \vec{x}')^{2}+ \cdots,\\
(\sigma^{2})^{(1)}|_{\eta_{\text{ret}}}&=& -2 \rho^{2} (\vec{n} \cdot \vec{x}')+ \rho r'^{2}+\cdots.
\eea
Therefore, to quadrupolar order, only $n=0$ and $n=1$ terms will contribute in Eq. \eqref{19X25.01}. Not considering the spatial integral for the time being, the temporal integral \eqref{19X25.01} becomes
\begin{align} 
  &\int_{0}^{\delta} d\eta' \bigg[V_{0}(\eta_{\text{ret}})\mu(\eta_{\text{ret}})+ \eta'\bigg(V_{0}^{(1)} \mu+V_{0} \mu^{(1)}\bigg)(\eta_{\text{ret}})\\
 &\hspace{4.5 cm}+V_{1}(\eta_{\text{ret}}) \sigma(\eta_{\text{ret}}) \mu(\eta_{\text{ret}})+ \eta'\bigg(\mu V_{1} \sigma^{(1)}\bigg)(\eta_{\text{ret}})
 \bigg]\\
& = \bigg(V_{0} \mu+ V_{1}\sigma \mu\bigg)(\eta_{\text{ret}}) \delta+\bigg(V_{0}^{(1)}\mu+V_{0} \mu^{(1)}+ \mu V_{1} \sigma^{(1)}\bigg)(\eta_{\text{ret}}) \frac{\delta^{2}}{2}\\ \nonumber
& = \bigg(V_{0} \mu+ V_{1}\sigma \mu\bigg)(\eta_{\text{ret}}) \bigg(\vec{n}\cdot \vec{x}'-\frac{r'^{2}-(\vec{n}\cdot \vec{x}')^{2}}{2\rho}\bigg)\\
&\hspace{5.0 cm}+\bigg(V_{0}^{(1)}\mu+V_{0} \mu^{(1)}+ \mu V_{1} \sigma^{(1)}\bigg)(\eta_{\text{ret}}) \frac{(\vec{n}\cdot \vec{x}')^{2}}{2}. 
 \end{align}
Using the definition of the moments in FLRW background up to the quadrupolar order, we have
\bea 
\int d^{3}x' x'_{k}\mu &=& \frac{1}{a}Q_{k} ,\label{eqQ}\\
\int d^{3}x' x'_{l}x'_{k} \mu^{(1)} &=& \int d^{3}x' a(t) x_{l}'x_{k}'\partial_{t}\mu= \frac{1}{a}(\partial_{{t}} Q_{kl}-2H Q_{kl}).
\eea
Using these relations, the complete expression \eqref{514} becomes 
\begin{align} 
T_2 = & \frac{\alpha(\alpha\pm 1)}{2} \frac{1}{\eta \eta_{\text{ret}}} \bigg(\frac{1}{a(\eta')} Q_{k}(\eta') n_{k}-\frac{1}{2\rho a^{2}(\eta')}Q_{kl}(\eta') (\delta_{kl}-n_{k} n_{l})\bigg)\bigg|_{\eta'=\eta_{\text{ret}}} \nonumber \\ \nonumber
&+\frac{1}{2} \frac{\alpha(\alpha\pm 1)}{2} \frac{1}{\eta \eta_{\text{ret}}}\bigg(\frac{1}{a(\eta')}(\partial_{t}-2H)Q_{kl}(\eta') n_{k}n_{l}\bigg)\bigg|_{\eta'=\eta_{\text{ret}}}\\ \label{19X25.07}
&+\frac{1}{2}\bigg(-\frac{\alpha(\alpha\pm 1)}{2 \eta \eta^{2}_{\text{ret}}}+\frac{\alpha(\alpha\pm 1)(\alpha(\alpha\pm 1)-2)}{8}\bigg(\frac{1}{\eta\eta_{\text{ret}}}\bigg)^{2} \rho\bigg)\bigg(\frac{1}{a^{2}(\eta')} Q_{kl}(\eta') n_{k} n_{l}
\bigg)\bigg|_{\eta'=\eta_{\text{ret}}}. 
\end{align}
The final tail integral is $T_1+T_2$. Though Eq. \eqref{19X25.07} is a part of the tail integral, the temporal integral can be performed exactly and it reduces to an instantenous term depending only upon retarded time.

\subsection{Light-cone integral up to quadrupolar order}

In this section, we complete the computation by deriving the explicit light-cone integral, i.e. the first term in Eq. \eqref{29IX25.01}, up to the quadrupolar order. Note that up to  quadrupolar order
\begin{align}  \nonumber
L_c := \int d^{3}x'\frac{\mu(\eta-|\bar{x}-\bar{x}'|,\bar{x}')}{|\bar{x}-\bar{x'}|}&= \int d^{3}x'\bigg(\frac{\mu+\vec{n}\cdot \vec{x}' \mu^{(1)}+\frac{1}{2}(\vec{n}\cdot \vec{x}')^{2}\mu^{(2)}+\cdots}{\rho} \\
&\hspace{-2cm}+\frac{\vec{n}\cdot \vec{x}'\mu+ \frac{1}{2}(3(\vec{n}\cdot \vec{x})^{2}-r'^{2})\mu^{(1)}+ \cdots}{\rho^2}+ \frac{(3(\vec{n}\cdot \vec{x})^{2}-r'^{2})\mu+ \cdots}{2\rho^{3}}
\bigg)\bigg|_{\eta_\text{ret}}.
\end{align}
Using the definition of the moments in FLRW background up to the quadrupolar order, we have Eq. \eqref{eqQ} as well as 
\begin{align}
\int d^{3}x' x'_{k}\mu^{(1)} &=\partial_{t} Q_{k}-HQ_{k},\\ \nonumber
\int d^{3}x' x'_{k}x'_{l}\mu^{(2)}&= \int d^{3} x' x'_{k}x'_{l}(a^{{2}}(t)\partial^{2}_{t}+Ha^{2}(t)\partial_{t}) \mu\\
&=\partial^{2}_{t} Q_{kl}-3H \partial_{t} Q_{kl}+2H^{2} Q_{kl}+2H^{2}(1+q) Q_{kl},
\end{align}
where we used $\partial_t H = -H^2 (1+q)$. Using these relations, we obtain
\begin{align} \nonumber
&L_c= \frac{1}
{\rho}\bigg(Q+n_{k}(\partial_{t} Q_{k}-H Q_{k})+\frac{1}{2} 
n_{k}n_{l}(\partial_{t}^{2} Q_{kl}\\ \nonumber
&-3H \partial_{t} Q_{kl}+2H^{2} Q_{kl}+2H^{2}(1+q)Q_{kl})
+\frac{1}{\rho a(\eta_{\text{ret}})}\big(n_{k} Q_{k}+\frac{1}{2}(3n_{k}n_{l}-\delta_{kl})(\partial_{t} Q_{kl}-2H Q_{kl})\big)\\ \label{29X25.01}
&+\frac{1}{2\rho^2 a^{2}(\eta_{\text{ret}})} (3n_{k}n_{l}-\delta_{kl})Q_{kl}
\bigg)\bigg\vert_{\eta_\text{ret}}.
\end{align}

\subsection{Tensor inhomogeneous solutions}

For power law cosmologies, the tensor part of the perturbations obeys Eq. \eqref{eqspowerlaw1}. It is identical to Eq. \eqref{9XI25.01} studied in Section \ref{sec:quad} for the $-$ branch up to the substitution 
\begin{align}
 \psi = \tilde \chi_{\langle ij \rangle}, \qquad \mu = 4G a T_{\langle ij \rangle}.    
\end{align}
From the analysis of the previous section, the solution up to quadrupolar order can be written as
\begin{align}\label{tensorsol00}
\tilde \chi_{\langle ij \rangle }(\eta,\rho,n_i) = L_{c\, \langle i j \rangle}    + T_{1\, \langle i j \rangle}+T_{2\, \langle i j \rangle}
\end{align}
where 
\begin{align} \nonumber
L_{c\, \langle ij \rangle} &= \frac{4G}
{\rho}\bigg(S_{\langle ij \rangle}+n_{k}(\partial_{t} S_{\langle ij \rangle|k}-H S_{\langle ij \rangle|k})+\frac{1}{2} 
n_{k}n_{l}(\partial_{t}^{2} S_{\langle ij \rangle|kl} -3H \partial_{t} S_{\langle ij \rangle|kl}+2H^{2} S_{\langle ij \rangle|kl}\nonumber \\ \nonumber
&+2H^{2}(1+q)S_{\langle ij \rangle|kl})
+\frac{1}{\rho a(\eta_{\text{ret}})}\bigg(n_{k} S_{\langle ij \rangle|k}+\frac{1}{2}(3n_{k}n_{l}-\delta_{kl})(\partial_{t} S_{\langle ij \rangle|kl}-2H S_{\langle ij \rangle|kl})\bigg)\\ 
&+\frac{1}{2\rho^2 a^{2}(\eta_{\text{ret}})} (3n_{k}n_{l}-\delta_{kl})S_{\langle ij \rangle|kl}
\bigg),\\
T_{1\, \langle i j \rangle}&=4G\int_{\eta_{i}}^{\eta -\rho} d\eta' \frac{\alpha(\alpha- 1)}{2\eta \eta'}\bigg[\ _{2}F_{1}\big(2 - \alpha , 1 + \alpha ; 2; \frac{p_\rho}{2}\big) S_{\langle ij \rangle}(\eta') \nonumber\\ \nonumber
&-\frac{1}{a(\eta')} \frac{\rho}{\eta \eta'} \frac{d}{dp_{\rho}} \bigg( \ _{2}F_{1}\big(2 - \alpha , 1 + \alpha ; 2; \frac{p_\rho}{2}\big) \bigg) n_{k} S_{\langle ij \rangle|k}(\eta')\\ \nonumber
&+ \frac{1}{a^{2}(\eta')} \frac{1}{2} \frac{1}{\eta \eta'}  \frac{d}{dp_{\rho}} \bigg( \ _{2}F_{1}\big(2 - \alpha , 1 + \alpha ; 2; \frac{p_\rho}{2}\big)\bigg) \delta_{kl} S_{\langle ij \rangle|kl}(\eta')\\ \label{19X25.8}
&+\frac{1}{a^{2}(\eta')} \frac{1}{2} \frac{\rho^2}{(\eta \eta')^{2}}  \frac{d^{2}}{d^{2}p_{\rho}} \bigg( \ _{2}F_{1}\big(2 - \alpha , 1 + \alpha ; 2; \frac{p_\rho}{2}\big)\bigg) n_{k} n_{l} S_{\langle ij \rangle|kl}(\eta')
\bigg],\\
T_{2\, \langle i j \rangle}&= 4G\bigg(\frac{\alpha(\alpha- 1)}{2} \frac{1}{\eta \eta_{\text{ret}}} \bigg(\frac{1}{a(\eta')} S_{\langle ij \rangle|k}(\eta') n_{k}-\frac{1}{2\rho a^{2}(\eta')}S_{\langle ij \rangle|kl}(\eta') (\delta_{kl}-n_{k} n_{l})\bigg)\bigg|_{\eta'=\eta_{\text{ret}}} \nonumber\\ \nonumber
&+\frac{1}{2} \frac{\alpha(\alpha - 1)}{2} \frac{1}{\eta \eta_{\text{ret}}}\bigg(\frac{1}{a(\eta')}(\partial_{t}-2H)S_{\langle ij \rangle|kl}(\eta') n_{k}n_{l}\bigg)\bigg|_{\eta'=\eta_{\text{ret}}}\\ \label{19X25.07bis}
&+\frac{1}{2}\bigg(-\frac{\alpha(\alpha - 1)}{2 \eta \eta^{2}_{\text{ret}}}+\frac{\alpha(\alpha - 1)(\alpha(\alpha - 1)-2)}{8}\bigg(\frac{1}{\eta\eta_{\text{ret}}}\bigg)^{2} \rho\bigg)\bigg(\frac{1}{a^{2}(\eta')} S_{\langle ij \rangle|kl}(\eta') n_{k} n_{l}
\bigg)\bigg|_{\eta'=\eta_{\text{ret}}}\bigg).
\end{align}
The conservation of the stress-energy tensor provides the explicit expressions of the higher multipoles in terms of the quadrupoles derived in Eqs. \eqref{decom}. 

For a generic value of $\alpha$, the tail integral $T_{1 \langle ij \rangle}$ scales as $\text{max}(\rho^{\alpha-2}, \rho^{-\alpha-1})$ in the limit $\rho \rightarrow\infty$ at fixed retarded time $\eta_\text{ret}=\eta-\rho$.

For the special cases of matter domination, $\alpha=2$, the hypergeometric function becomes the identity, and the tail integral $T_{1 \langle ij \rangle}$ simplifies to 
\bea
T_{1\, \langle i j \rangle}&=4G\int_{\eta_{i}}^{\eta -\rho}  \frac{d\eta'}{\eta \eta'} S_{\langle ij \rangle}(\eta') .
\eea
Moreover, the linear in $\rho$ term in the $T_{2\, \langle i j \rangle}$ term exactly vanishes. Therefore, all $\rho$ dependency in $\chi_{\langle ij \rangle}$ exactly cancels in the tail integrals at fixed retarded time $\eta_\text{ret}=\eta-\rho$. We can further use the conservation of the stress-energy tensor \eqref{eqn37} for $L=\emptyset$, $S_{ij}=-\partial_t P_{i\vert j}$ and rewrite the integral as 
\begin{align}
T_{1\, \langle i j \rangle}&=\frac{4G}{\alpha \eta} \int_{\eta_{i}}^{\eta_\text{ret}} d\eta' \frac{\partial_{\eta'}a}{a(\eta')} S_{\langle ij \rangle}(\eta')= \frac{4G}{\alpha \eta} \int_{\eta_{i}}^{\eta_\text{ret}} d\eta' a(\eta')H(\eta') S_{\langle ij \rangle}(\eta')\nonumber \\ 
& =  -\frac{4G}{\alpha \eta} \int_{\eta_{i}}^{\eta_\text{ret}} d\eta' H(\eta') \partial_{\eta'} P_{\langle i \vert j \rangle}(\eta') \nonumber \\ 
&= - \frac{4G}{\alpha \eta} \left[ H P_{\langle i \vert j \rangle}\right]_{\eta_{i}}^{\eta_\text{ret}} + \frac{4G}{\alpha \eta} \int_{\eta_{i}}^{\eta_\text{ret}} d\eta' \partial_{\eta'} H(\eta')  P_{\langle i \vert j \rangle}(\eta')\\
&= - \frac{4G}{\alpha \eta} \left[ H P_{\langle i \vert j \rangle}\right]_{\eta_{i}}^{\eta_\text{ret}} - \frac{4G}{\alpha \eta} \int_{t_{i}}^{t_\text{ret}} dt' H^{2} (1+q) P_{\langle i \vert j \rangle}(t'). \label{tailtermP}
\end{align}
Due to the fact that the Hubble ``constant'' is varying in matter dominated cosmologies, the tail integral is non-trivial. Gravitational waves propagate inside the light-cone in these spacetimes. This behavior differs from gravitational waves in de Sitter spacetime where the tail integral reduced to an instantenous term and a boundary term at the past cosmological horizon \cite{Compere:2023ktn}. 

Let us now depart from power law cosmologies, and briefly discuss the case of de Sitter. We obtained that $\tilde \chi_{ij}$ obeys the decoupled equation \eqref{dSeq}. This again matches with Eq. \eqref{9XI25.01} up to the substitution 
\begin{align}
 \psi = \tilde \chi_{ij}, \qquad \mu = 4G a T_{ij}, \qquad \alpha=-1. 
\end{align}
Our formalism therefore provides on-the-fly the solution for de Sitter tensor perturbations and the scalar perturbation $\chi_{ii}$ as well. The tensor solution is exactly \eqref{tensorsol00} with $\alpha=-1$ and $a=-1/(H \eta)$. After analysis, we obtain the final answer 
\begin{align} \nonumber
{\chi_{ij}}:=a^{-1}\tilde \chi_{ij} &=4GH^{2}P_{(i|j)}(-\infty)+4GH^{2}\bigg(-P_{(i|j)}+n_{k}S_{ij|k}+\frac{1}{2} n_{k}n_{l} \partial_{t}S_{ij|kl}-\frac{H}{2}S_{ij|kk}\bigg)\\
\nonumber
& +4G\bigg(\frac{1}{a(\eta_{\text{ret}}) \rho}-H\bigg)\bigg( S_{ij}+n_k\partial_t S_{ij \vert k}+\frac{n_k n_l}{2} \partial_{t}^{2} S_{ij|kl}-\frac{H}{2} \partial_{t} S_{ij|kk}\bigg)\nonumber \\
& +4G\bigg(\frac{1}{a(\eta_{\text{ret}}) \rho}-H\bigg)^{2}\bigg(n_{k}S_{ij|k}+\frac{1}{2}(3n_{k}n_{l}-\delta_{kl})\partial_{t} S_{ij|kl}\bigg)\nonumber \\ 
&+2 G\bigg(\frac{1}{a(\eta_{\text{ret}})\rho} -H\bigg)^{3} (3n_{k}n_{l}-\delta_{kl}) S_{ij|kl}
.\label{finalchijbis}
\end{align}
For de Sitter the tail integral can be performed exactly and it reduces to an instantaneous term and a boundary term on the initial time slice $\eta=\eta_i$ as $q=-1$ and the second term in Eq. \eqref{tailtermP} vanishes. 
The solution \eqref{finalchijbis} exactly matches with the de Sitter solution obtained in  Eq. (2.92) of \cite{Compere:2023ktn} after neglecting the boundary terms at the past cosmological horizon. This provides a non-trivial cross-check of our solution \eqref{tensorsol00}.

\subsection {Vector inhomogeneous solutions}

For power law cosmologies, the vector part of the perturbations obeys Eq. 
\eqref{eqspowerlaw2}. It is identical to Eq. \eqref{9XI25.01} studied in 
Section \ref{sec:quad} for the $+$ branch with the substitution 
\begin{align}
 \psi = \tilde \chi_{0i}, \qquad \mu = 4G a T_{0i}.    
\end{align}
From the analysis of the previous section, the solution up to quadrupolar order can be written as
\begin{align}\label{tensorsol}
\tilde \chi_{0i}(\eta,\rho,n_i) = L_{c\, i}    + T_{1\, i}+T_{2\, i}
\end{align}
where 
\begin{align} \nonumber
L_{c\, i} &=\frac{4G}
{\rho}\bigg(P_{i}+n_{k}(\partial_{t} P_{i|k}-H P_{i|k})+\frac{1}{2} 
n_{k}n_{l}(\partial_{t}^{2} P_{i|kl}
-3H \partial_{t} P_{i|kl}+2H^{2} P_{i|kl}
\\ \nonumber &
+2H^{2}(1+q)P_{i|kl})
+\frac{1}{\rho a(\eta_{\text{ret}})}\bigg(n_{k} P_{i|k}+\frac{1}{2}(3n_{k}n_{l}-\delta_{kl})(\partial_{t} P_{i|kl}-2H P_{i|kl})\bigg)\\
&+\frac{1}{2\rho^2 a^{2}(\eta_{\text{ret}})} (3n_{k}n_{l}-\delta_{kl})P_{i|kl}
\bigg),\\ \nonumber
T_{1\, i} &=4G\int_{\eta_{i}}^{\eta -\rho} d\eta' \frac{\alpha(\alpha+ 1)}{2\eta \eta'}\bigg[\ _{2}F_{1}\big(2 + \alpha , 1 - \alpha ; 2; \frac{p_\rho}{2}\big) P_{i}(\eta')\\ \nonumber
&-\frac{1}{a(\eta')} \frac{\rho}{\eta \eta'} \frac{d}{dp_{\rho}} \bigg( \ _{2}F_{1}\big(2 + \alpha , 1 - \alpha ; 2; \frac{p_\rho}{2}\big)\bigg) n_{k} P_{i|k}(\eta')\\ \nonumber
&+ \frac{1}{a^{2}(\eta')} \frac{1}{2} \frac{1}{\eta \eta'}  \frac{d}{dp_{\rho}} \bigg( \ _{2}F_{1}\big(2 + \alpha , 1 - \alpha ; 2; \frac{p_\rho}{2}\big)\bigg) \delta_{kl} P_{i|kl}(\eta')\\ 
&+\frac{1}{a^{2}(\eta')} \frac{1}{2} \frac{\rho^2}{(\eta \eta')^{2}}  \frac{d^{2}}{d^{2}p_{\rho}} \bigg( \ _{2}F_{1}\big(2 + \alpha , 1 - \alpha ; 2; \frac{p_\rho}{2}\big)\bigg) n_{k} n_{l} P_{i|kl}(\eta')
\bigg],\\ \nonumber
T_{2 \, i} & = 4G\bigg(\frac{\alpha(\alpha+ 1)}{2} \frac{1}{\eta \eta_{\text{ret}}} \bigg(\frac{1}{a(\eta')} P_{i|k}(\eta') n_{k}-\frac{1}{2\rho a^{2}(\eta')}P_{i|kl}(\eta') (\delta_{kl}-n_{k} n_{l})\bigg)\bigg|_{\eta'=\eta_{\text{ret}}} \\ \nonumber
&+\frac{1}{2} \frac{\alpha(\alpha+ 1)}{2} \frac{1}{\eta \eta_{\text{ret}}}\bigg(\frac{1}{a(\eta')}(\partial_{t}-2H)P_{i|kl}(\eta') n_{k}n_{l}\bigg)\bigg|_{\eta'=\eta_{\text{ret}}}\\ 
&+\frac{1}{2}\bigg(-\frac{\alpha(\alpha+ 1)}{2 \eta \eta^{2}_{\text{ret}}}+\frac{\alpha(\alpha+ 1)(\alpha(\alpha+ 1)-2)}{8}\bigg(\frac{1}{\eta\eta_{\text{ret}}}\bigg)^{2} \rho\bigg)\bigg(\frac{1}{a^{2}(\eta')} P_{i|kl}(\eta') n_{k} n_{l}
\bigg)\bigg|_{\eta'=\eta_{\text{ret}}}\bigg). 
\end{align}
Using the decomposition in Eq. \eqref{decom}, we can extract the odd and even parity modes up to quadrupolar order.

\subsection{Scalar inhomogeneous solutions}

For power law cosmologies, the scalar perturbations obey Eqs. \eqref{eqspowerlaw3}, \eqref{eqspowerlaw4}. These two equations are identical to the two $\pm$ branches of Eq. \eqref{9XI25.01} studied in Section \ref{sec:quad} up to the substitution 
\begin{align}
 \psi = \tilde \chi_{\pm}, \qquad \mu=4G a \delta \tilde T_\pm,    
\end{align}
where we defined for convenience the two scalar sectors: 
\begin{align}
&\tilde \chi_+ \equiv \tilde \chi_{00}+\tilde \chi_{ii}, \qquad \delta \tilde T_+ \equiv \delta \tilde T_{00}+\delta \tilde T_{ii}, \\
&\tilde \chi_- \equiv \tilde \chi_{ii},\quad \;\;\quad \qquad \delta \tilde T_- \equiv \delta \tilde T_{ii}. 
\end{align}

From the analysis of the previous section, the solution up to quadrupolar order can be written as
\begin{align}\label{tensorsolLLL}
\tilde \chi_{\pm}(\eta,\rho,n_i) = L_{c\, \pm}    + T_{1\, \pm}+T_{2\, \pm}
\end{align}
where 
\begin{align} \nonumber
L_{c\, \pm} &=\frac{4G}
{\rho}\bigg({Q}_{\pm}+n_{k}(\partial_{t} {Q}_{\pm k}-H {Q}_{\pm k})+\frac{1}{2} 
n_{k}n_{l}\big(\partial_{t}^{2} {Q}_{\pm kl}
-3H \partial_{t} {Q}_{\pm kl}+2H^{2} {Q}_{\pm kl}
\\ \nonumber &
+2H^{2}(1+q){Q}_{\pm kl}\big)
+\frac{1}{\rho a(\eta_{\text{ret}})}\bigg(n_{k} {Q}_{\pm k}+\frac{1}{2}(3n_{k}n_{l}-\delta_{kl})(\partial_{t}{Q}_{\pm kl}-2H {Q}_{\pm kl})\bigg)\\
&+\frac{1}{2\rho^2 a^{2}(\eta_{\text{ret}})} (3n_{k}n_{l}-\delta_{kl}){Q}_{\pm kl}
\bigg),\\ \nonumber
T_{1\, \pm} &=4G\int_{\eta_{i}}^{\eta -\rho} d\eta' 
\frac{\alpha(\alpha\pm 1)}{2\eta \eta'}\bigg[ \ _{2}F_{1}\big(2 \pm \alpha , 1 \mp \alpha ; 2; \frac{p_\rho}{2}\big) Q_{\pm}(\eta')\\ \nonumber
&-\frac{1}{a(\eta')} \frac{\rho}{\eta \eta'} \frac{d}
{dp_{\rho}} \bigg( \ _{2}F_{1}\big(2 \pm \alpha , 1 \mp \alpha ; 2; \frac{p_\rho}{2}\big)\bigg) n_{k} Q_{\pm k}(\eta')\\ \nonumber
&+ \frac{1}{a^{2}(\eta')} \frac{1}{2} \frac{1}{\eta \eta'}  
\frac{d}{dp_{\rho}} \bigg(  \ _{2}F_{1}\big(2 \pm \alpha , 1 \mp \alpha ; 2; \frac{p_\rho}{2}\big)\bigg) \delta_{kl} Q_{\pm kl}(\eta')\\ 
&+\frac{1}{a^{2}(\eta')} \frac{1}{2} \frac{\rho^2}{(\eta 
\eta')^{2}}  \frac{d^{2}}{d^{2}p_{\rho}} \bigg(  \ _{2}F_{1}\big(2 \pm \alpha , 1 \mp \alpha ; 2; \frac{p_\rho}{2}\big) \bigg) n_{k} n_{l} Q_{\pm kl}
(\eta')
\bigg],\\ \nonumber
T_{2 \, \pm} & = 4G\bigg(\frac{\alpha(\alpha\pm 1)}{2} \frac{1}{\eta 
\eta_{\text{ret}}} \bigg(\frac{1}{a(\eta')} Q_{\pm k}(\eta') 
n_{k}-\frac{1}{2\rho a^{2}(\eta')}Q_{\pm kl}(\eta') 
(\delta_{kl}-n_{k} 
n_{l})\bigg)\bigg|_{\eta'=\eta_{\text{ret}}} \\ \nonumber
&+\frac{1}{2} \frac{\alpha(\alpha\pm 1)}{2} \frac{1}{\eta 
\eta_{\text{ret}}}\bigg(\frac{1}{a(\eta')}
(\partial_{t}-2H)Q_{ \pm kl}(\eta') 
n_{k}n_{l}\bigg)\bigg|_{\eta'=\eta_{\text{ret}}}\\ 
&+\frac{1}{2}\bigg(-\frac{\alpha(\alpha\pm 1)}{2 \eta 
\eta^{2}_{\text{ret}}}+\frac{\alpha(\alpha\pm 1)(\alpha(\alpha\pm 1)-2)}
{8}\bigg(\frac{1}{\eta\eta_{\text{ret}}}\bigg)^{2} 
\rho\bigg)\bigg(\frac{1}{a^{2}(\eta')} Q_{\pm kl}(\eta') n_{k} n_{l}
\bigg)\bigg|_{\eta'=\eta_{\text{ret}}}\bigg),
\end{align}
where $Q_{\pm L}$ are the multipole moments associated with $\delta \tilde T_\pm$, i.e.
\bea 
Q_{\pm L}=\int d^{3}x a^{\ell +1} \delta \tilde T_\pm x_{L}.
\eea


\section{Conclusion}
\label{sec:ccl}

We found a gauge in which perturbations around the FRLW background totally decouple, without requiring the use of a SVT decomposition. We found that tensor, vector and scalar perturbations (in the sense of a symmetric traceless decomposition) are governed by Minkowski wave operators with two distinct potentials. One feature of our formalism is the identification of a non-conservative term in the conservation law of the stress-energy tensor proportional to the time derivative of the sum of the background fluid pressure and energy. Because of that term, one cannot define a compactly supported perturbed stress-energy tensor, but only a nearly compact stress-energy tensor where the energy density perturbation is still non-compact. This feature departs from perturbation theory around de Sitter spacetime. 

Assuming power law cosmologies, we derived the Green's functions for the two relevant potentials, which contain a light-cone part and a tail part, in terms of a hypergeometric function. We found that it matches a Green's function derived by Chu \cite{Chu:2016qxp,Chu:2016ngc}, and compared our results with other proposals \cite{Waylen:1978dx,a2540274-da7b-3e43-b5e5-9617401299cd,Haas:2004kw}.  
Using the consistent quadrupolar truncation defined in \cite{Compere:2023ktn}, we derived the metric perturbation up to quadrupolar order for an arbitrary nearly-compact source. This metric perturbation not only includes the propagating GW but also the scalar potentials that are generated by the nearly-compact source. We emphasize that our formalism applies beyond the geometric optics approximation and is therefore relevant for the study of perturbations produced by primordial sources.

\section*{Acknowledgements}
We thank Abraham Harte for his useful comments. G.C. is Research Director of the F.R.S.-FNRS. The work of JH is supported in part by MSCA Fellowships CZ - UK2 $(\mbox{reg. n. CZ}.02$ $.01.01/00/22\_010/0008115)$
from the Programme Johannes Amos Comenius co-funded by the European Union. 

\bibliography{FLRW}
\end{document}